\documentclass{aa}  

\usepackage{natbib}
\bibpunct{(}{)}{;}{a}{}{,}

\usepackage{graphicx}
\usepackage{txfonts}
\newcommand{\vai}{c_{\mathrm{A,i}}}

\newcommand{\pd}{\partial}

\newcommand{\rhoe}{\rho_{\mathrm{e}}}
\newcommand{\rhoi}{\rho_{\mathrm{i}}}
\newcommand{\rhon}{\rho_{\mathrm{n}}}
\newcommand{\ain}{\alpha_{\mathrm{in}}}
\newcommand{\aen}{\alpha_{\mathrm{en}}}
\newcommand{\nuni}{\nu_{\mathrm{ni}}}
\newcommand{\nuin}{\nu_{\mathrm{in}}}
\newcommand{\vi}{{\bf v}_{\mathrm{i}}}
\newcommand{\vn}{{\bf v}_{\mathrm{n}}}
\newcommand{\ve}{{\bf v}_{\mathrm{e}}}

\newcommand{\aie}{\alpha_{\mathrm{ie}}}
\newcommand{\nuei}{\nu_{\mathrm{ei}}}

\newcommand{\oci}{\Omega_{\mathrm{i}}}

\newcommand{\oce}{\Omega_{\mathrm{e}}}
\newcommand{\nuen}{\nu_{\mathrm{en}}}

\begin{document}

 \title{Overdamped  Alfv\'en waves due to ion-neutral\\ collisions in the solar chromosphere}

\titlerunning{Overdamped  Alfv\'en waves}

   \author{R. Soler\inst{1} \and J.~L. Ballester\inst{1}
          \and
          T.~V. Zaqarashvili\inst{2,3}}

   \institute{Departament de F\'isica, Universitat de les Illes Balears,
              E-07122 Palma de Mallorca, Spain\\
              \email{[roberto.soler;joseluis.ballester]@uib.es}
         \and
             Space Research Institute, Austrian Academy of Sciences, Schmiedlstrasse 6, 8042 Graz, Austria
             \and
             Abastumani Astrophysical Observatory at Ilia State University, 3/5 Cholokashvili Avenue, 0162 Tbilisi, Georgia \\
               \email{teimuraz.zaqarashvili@oeaw.ac.at}
             }

   \date{Received XXX; accepted XXX}

 
  \abstract
   {Alfv\'enic waves are ubiquitous in the solar atmosphere and their dissipation may play an important role in atmospheric heating. In the partially ionized solar chromosphere, collisions between ions and neutrals are an efficient dissipative mechanism for Alfv\'en waves with frequencies near the ion-neutral collision frequency. The collision frequency is proportional to the ion-neutral collision cross section for momentum transfer.   Here, we investigate  Alfv\'en wave damping as a function of height in a simplified chromospheric model and compare the results for two  sets of collision cross sections, namely those of the classic hard-sphere model and those based on recent quantum-mechanical computations. We find important differences between the results for the two sets of cross sections. There is a critical interval of wavelengths for which impulsively excited Alfv\'en waves are overdamped as a result of the  strong ion-neutral dissipation. The critical wavelengths are in the range from 1~km to 50~km for the hard-sphere cross sections, and from 1~m to 1~km for the quantum-mechanical cross sections. Equivalently, for periodically driven Alfv\'en waves there is an optimal frequency for which the damping is most effective. The optimal  frequency varies from 1~Hz to $10^2$~Hz for the hard-sphere cross sections, and from $10^2$~Hz to $10^4$~Hz for the quantum-mechanical cross sections.  Future observations at sufficiently high spatial or temporal resolution may show the importance of high-frequency Alfv\'en waves for chromospheric heating. For instance,  the {\em Atacama Large Millimeter/submillimeter Array} (ALMA) may be able to detect the critical wavelengths and optimal frequencies  and so to test the effective collision cross section in the chromospheric plasma. }

   \keywords{Magnetohydrodynamics (MHD) --- Sun: atmosphere --- Sun: chromosphere --- Sun: oscillations --- Waves}

   \maketitle
%

\section{Introduction}

Recent high-resolution observations indicate that Alfv\'en waves are ubiquitous throughout the solar atmosphere.  In particular, Alfv\'enic waves have been detected propagating in the chromosphere \citep[e.g.,][]{kukhianidze2006,zaqarashvili2007,depontieu2007,okamoto2011,kuridze2012,depontieu2012,morton2013,morton2014}. It is believed that energy transport by Alfv\'enic-type waves and its dissipation  may play a relevant role in the heating and energy balance in the  atmospheric plasma \citep[see, e.g.,][]{erdelyi2007,cargill2011,mcintosh2011,hahn2014}. Both observations and theoretical aspects of Alfv\'en wave propagation have been reviewed by  \citet{zaqarashvili2009} and \citet{mathioudakis2013}. 

The relatively cool temperature in the chromosphere causes the plasma to be partially ionized, with a predominance of neutrals at low heights in the chromosphere. In this context, ion-neutral collisions have been invoked as a viable energy dissipation mechanism for Alfv\'en waves by, e.g., \citet{depontieu2001,khodachenko2004,leake2005,russell2013}, among others. Estimations of the heating rate due to Alfv\'en waves damped by ion-neutral collisions computed by \citet{song2011} and \citet{goodman2011}  indicate that this mechanism can generate sufficient heat to compensate the radiative losses at low altitudes in the solar atmosphere. For driven waves propagating from the photosphere to the corona, numerical simulations by \citet{tu2013} show that the wave energy flux transmitted to the corona is at least one order of magnitude smaller than that of the driving source due to the reflection and strong damping in the chromosphere. Therefore, partial ionization seems to  be a crucial ingredient to correctly understand the processes of plasma heating and energy transport in the chromosphere \citep[e.g.,][]{khomenko2012,martinez2012,leake2013}, including those processes that involve propagation and dissipation of Alfv\'en waves.  

Damping of Alfv\'en waves due to ion-neutral collisions in the solar chromosphere is usually investigated in the single-fluid approximation, which assumes a strong coupling between the various species in the plasma \citep[e.g.,][]{depontieu2001,khodachenko2004}. Theoretical studies demonstrate that the damping is most efficient when the frequency of the wave and the ion-neutral collision frequency are of the same order of magnitude \citep[see][]{zaqarashvili2011,soler2013}. However, the single-fluid approximation breaks down for wave frequencies near the ion-neutral collision frequency because the dynamics of ions and neutrals decouple. Ions and neutrals have to be considered as two separate fluids for those high frequencies. Multi-fluid approaches are therefore necessary to correctly investigate high-frequency Alfv\'en waves in the  chromosphere \citep{zaqarashvili2011,soler2013,soler2013ma,soler2013tube}. In a general sense, both temporal and spatial scales should be taken into account to assess the applicability of the single-fluid approximation. For the waves of interest in this paper, however, it is enough to consider the criterion based on the wave frequency alone, since frequencies and wavelengths are intrinsically related. High frequencies are equivalent to short wavelengths, and vice versa.  

The present paper deals with the theoretical study of Alfv\'en wave damping in the partially ionized chromosphere using the multi-fluid theory. The multi-fluid equations used here are rather general and contain a number of physical effects that were not considered in previous works  \cite[e.g.,][]{soler2013}. For instance, we consistently take into account ion-neutral collisions, electron-neutral collisions, electron inertia, viscosity, Ohm's magnetic diffusion, and Hall's current. This additional physics provides a more realistic representation of the chromospheric plasma. This investigation is also related to the recent work by \citet{vranjes2014} although, as we explain later, our results and those of \citet{vranjes2014} do not agree regarding the existence of a strict frequency cut-off for Alfv\'en waves.

The three main goals of this work are summarized as follows. (1) We discuss a fact usually ignored in the literature concerning the applicability of the  multi-fluid theory. The correct study of  waves with frequencies near the ion-neutral collision frequency cannot be done in the usual framework of the multi-fluid theory in the low chromosphere \citep[see][]{vranjes2013}.  Fluid theory is only applicable at sufficiently large heights (see Section~\ref{sec:chromos} for details). (2) We  show that the ion-neutral collision cross section is a very important parameter for determining the wavelengths and wave frequencies that are most efficiently damped. Different values of the  ion-neutral collision cross section and frequency can be found in the literature  \citep[see, e.g., the various expressions of the collision frequency given in the Appendix of][]{depontieu2001}. An important goal of the present work is to determine whether the use of the more accurate cross sections recently proposed by \citet{vranjes2013}  modifies substantially the results obtained with the classical  cross sections or, on the contrary, the numerical value of the cross sections plays a minor role.   (3) In connection to the previous point, we suggest that future observations with instruments operating at very high temporal and spatial resolutions like, e.g., the {\em Atacama Large Millimeter/submillimeter Array} (ALMA), may be crucial to understand how these waves actually propagate and damp in the chromosphere and thus how they contribute to plasma heating \citep[see][]{karlicky2011}.

This paper is organized as follows. Section~\ref{sec:chromos} contains the description of the chromospheric model adopted in this work. Discussions about the applicability of the fluid theory and the importance of the ion-neutral collision cross section are also given. The basic multi-fluid equations and the dispersion relation of Alfv\'en waves are given in Section~\ref{sec:twofluid}.  Then, Section~\ref{sec:damping} contains computations of the quality factor for Alfv\'en wave damping as a function of height in the chromosphere. Both the impusive driver and the periodic driver scenarios are analyzed. Finally, the discussion of the results and their observational implications are included in Section~\ref{sec:dis}.


\section{Chromospheric model}

\label{sec:chromos}

\subsection{Variation with height of physical parameters}

In this work, we adopt a simplified one-dimensional model for the chromosphere. We treat the chromospheric medium as a  partially ionized hydrogen plasma composed of ions (protons), electrons, and neutrals (hydrogen atoms). The influence of heavier species, specially that of helium, is ignored. The specific effect of helium on the damping of Alfv\'en waves was studied by \citet{soler2010}, \citet{zaqarashvili2011helium}, and \citet{zaqarashvili2013}. The reader is refereed to these previous works for details.  The subscripts `i', `e', and `n'  denote ions, electrons, and neutrals, respectively, while the subscript `$\beta$' is used to refer to a unspecified species.  MKS units are used throughout this paper.

\begin{figure}
   \centering
  \includegraphics[width=0.95\columnwidth]{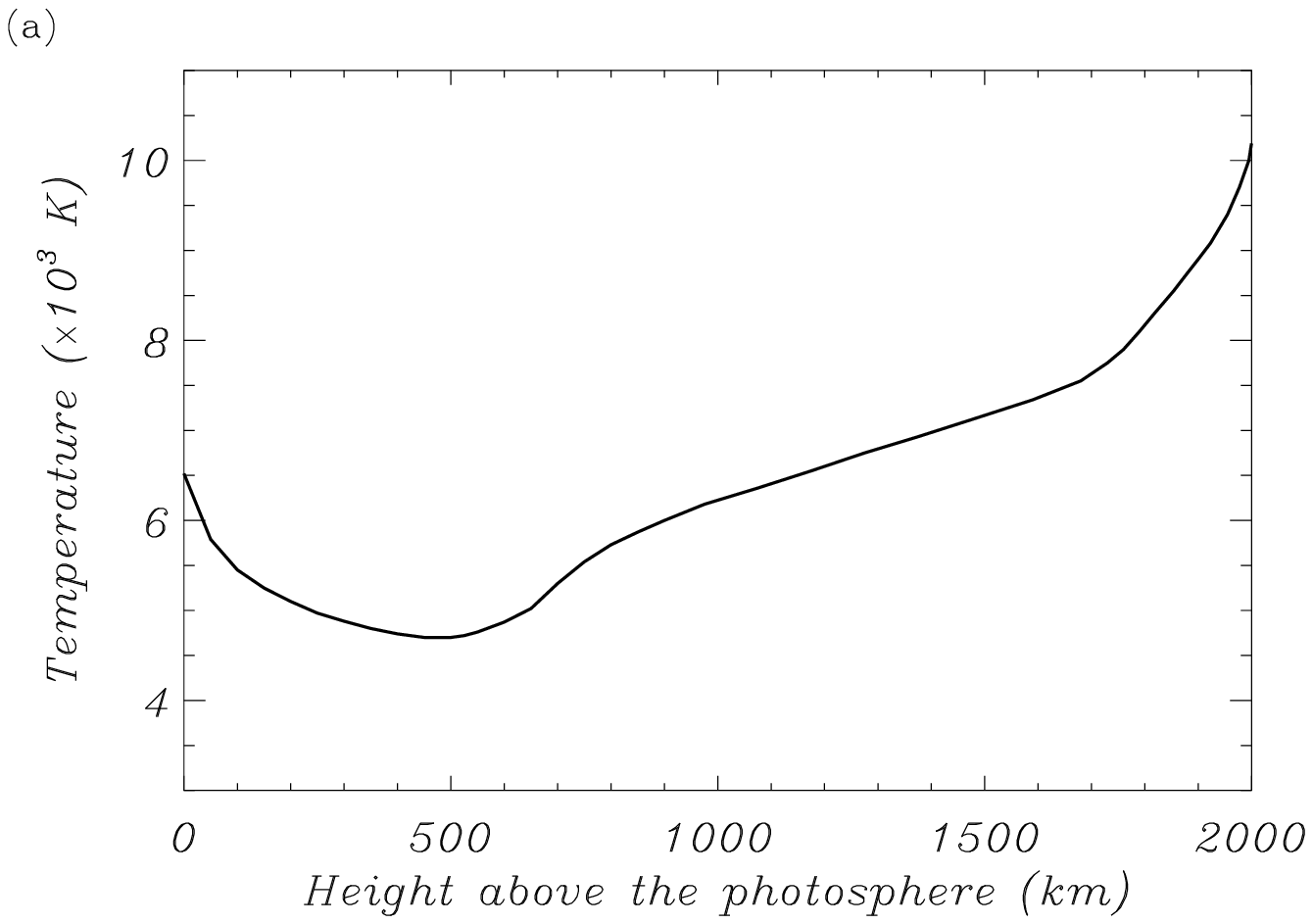}
    \includegraphics[width=0.95\columnwidth]{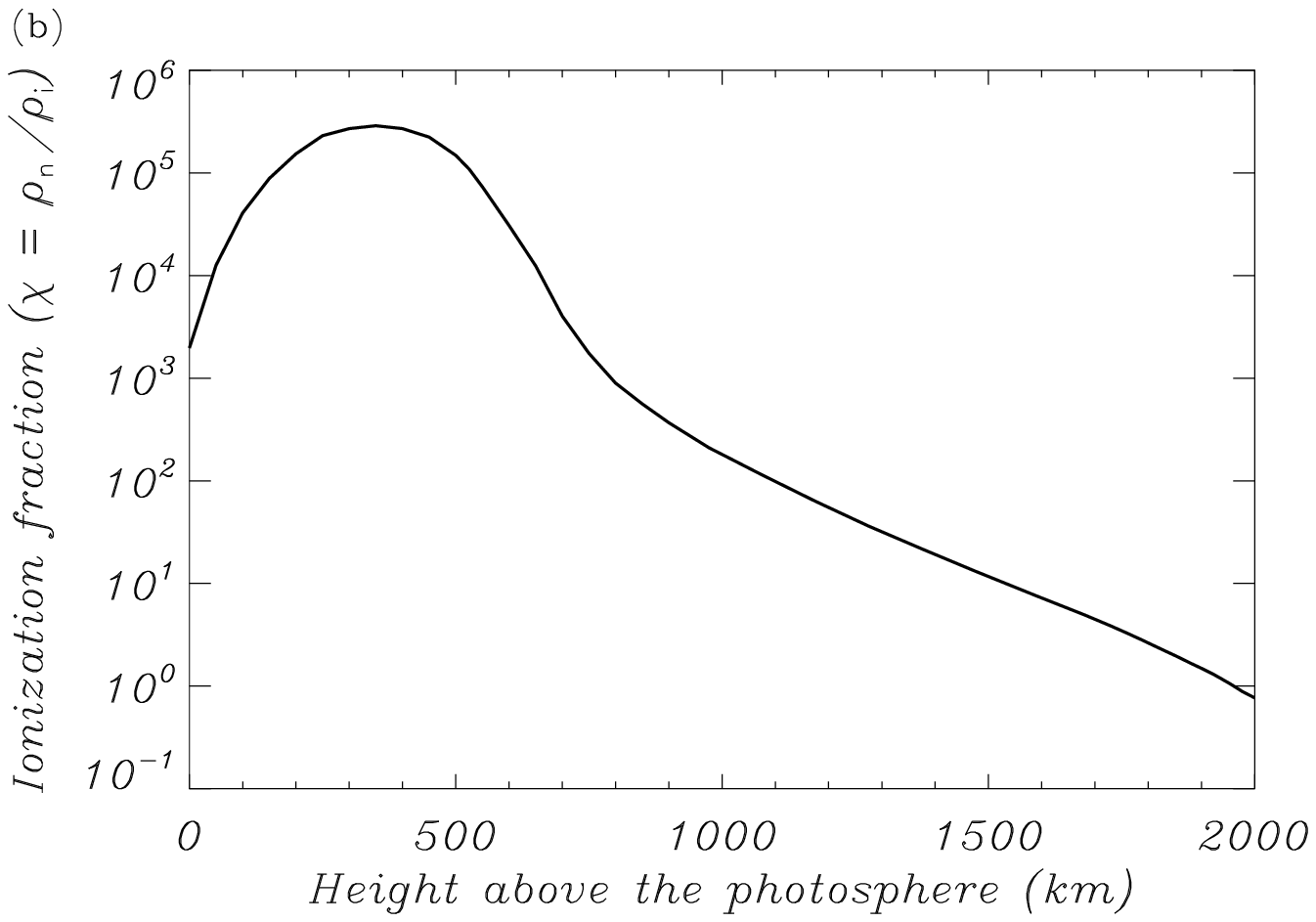}
  \includegraphics[width=0.95\columnwidth]{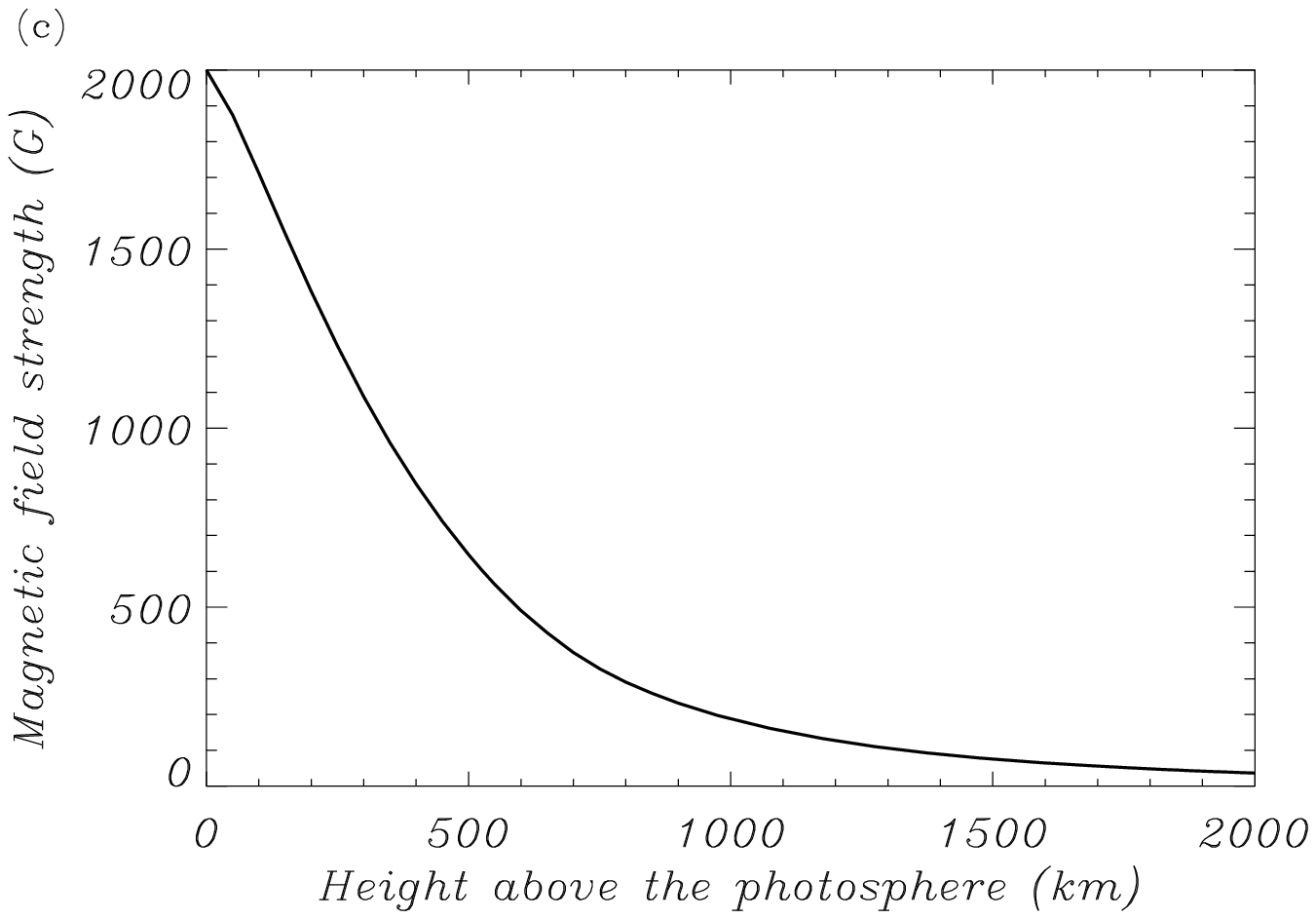}
   \caption{Dependence on height  above the solar photosphere of the (a) plasma temperature and (b) hydrogen ionization fraction, $\chi = \rhon/\rhoi$, according to the FAL93-F model. (c) Magnetic field strength as a function of height according to the semi-empirical relation of  Equation~(\ref{eq:magfield}). We note the logarithmic scale in the vertical axis of panel (b).}
              \label{fig:falc1}%
    \end{figure}

We use the chromospheric bright region model F of \citet{fontela1993}, hereafter the FAL93-F model, to account for the variation of the physical conditions with height, that here corresponds to the $z$-direction in Cartesian coordinates. Figures~\ref{fig:falc1}(a) and (b) show the dependence on height above the photosphere, $h$, of the plasma temperature, $T$, and the ionization fraction, $\chi = \rhon / \rhoi$, where $\rhoi$ and $\rhon$ are the ion and neutral densities, respectively. The photosphere corresponds to $h=0$. The plasma is permeated by a vertical magnetic field, namely ${\bf B} = B \hat{e}_z$, with $B$ the magnetic field strength.  The dependence of the magnetic field strength on height is taken after the semi-empirical formula by \citet{leake2006}, namely 
\begin{equation}
B = B_{\rm ph} \left( \frac{\rho}{\rho_{\rm ph}} \right)^{0.3}, \label{eq:magfield}
\end{equation}
where $\rho$ is the total mass density, whose dependence on height is prescribed by the FAL93-F model, $\rho_{\rm ph}$ is the total density at the photospheric level (also taken from the FAL93-F model), and $B_{\rm ph}$ is the photospheric magnetic field strength. Equation~(\ref{eq:magfield}) roughly represents the magnetic field strength in an intense  magnetic flux tube expanding  with height. We consider an intense magnetic element at the photospheric level and use $B_{\rm ph} = 2$~kG. Figure~\ref{fig:falc1}(c) shows the dependence on height of the magnetic field strength. In this model the magnetic field strength decreases with height and is $B\approx 190$~G at 1,000~km  and  $B\approx 35$~G at 2,000~km above the photosphere.  The largest variation of $B$ takes place for $h \lesssim$~1,000~km,  that represents the strong expansion of the magnetic field at low heights in the chromosphere. On the contrary, for $h \gtrsim$~1,000~km the variation of $B$ is much less important and $B$ becomes almost constant at large heights. We note that a horizontal component of the magnetic field should be included in order to satisfy the divergence-free condition.  However, this geometrical effect would complicate matters substantially. For this reason, we restrict ourselves to a purely vertical magnetic field and neglect the horizontal component. This is approximately valid near the axis of the flux tube, where the horizontal component is much smaller than the vertical component. 
 
It should be noted that the simplified model used here represents a static, gravitationally stratified chromosphere. Therefore, the model misses part of the highly dynamical behavior of the chromospheric medium seen in both high-resolution observations and numerical simulations \citep[e.g.,][]{martinez2012}.  A time-dependent background is necessary to account for all the fast, time-varying dynamics of the chromospheric plasma. This is, however, far beyond the purpose of the present work.

\subsection{Collision frequencies and applicability of the fluid theory} 
 
We use the three-fluid theory for a partially ionized plasma, in which ions, electrons, and neutrals are considered as separate fluids. The three fluids exchange momentum by means of particle collisions. The collision frequency of species $\beta$ with species $\beta'$, namely $\nu_{\beta\beta'}$, is defined by means of the symmetric friction coefficient $\alpha_{\beta \beta'}$, so that $\nu_{\beta\beta'} = \alpha_{\beta \beta'} / m_\beta n_\beta $, where $m_\beta$ and $n_\beta$ are the mass and number density of species $\beta$, respectively. The friction coefficient between two charged species, namely $\beta$ and $\beta'$, is \citep[e.g.,][]{spitzer1962,brag}
\begin{equation}
\alpha_{\beta\beta'} = \frac{n_{\beta} n_{\beta'}Z_\beta^2 Z_{\beta'}^2 e^4 \ln \Lambda_{\beta\beta'}}{6\pi\sqrt{2\pi} \epsilon_0^2 m_{\beta \beta'} \left( k_{\rm B} T_\beta / m_\beta +  k_{\rm B} T_{\beta'} / m_{\beta'} \right)^{3/2}}, \label{eq:fric}
\end{equation}
while the friction coefficient between a charged or neutral species, $\beta$, and a neutral species, n, is \citep[e.g.,][]{brag,chapman1970}
\begin{equation}
\alpha_{\beta n} =  n_{\beta} n_{\rm n} m_{\beta \rm n} \left[\frac{8 k_{\rm B}}{\pi } \left( \frac{T_\beta}{m_\beta} + \frac{T_{\rm n}}{m_{\rm n}} \right)\right]^{1/2}\sigma_{\rm \beta n}. \label{eq:fricneu}
\end{equation}
In these expressions,  $m_{\beta\beta'} = m_\beta m_{\beta'}/\left( m_\beta + m_{\beta'} \right)$ is the reduced mass, $T_\beta$ is the temperature of species $\beta$, $e$ is the electron charge, $k_{\rm B}$ is Boltzmann's constant,   $\epsilon_0$ is the permittivity of free space, $Z_\beta$ is the sign of the electric charge ($Z_{\rm i} = 1$ and $Z_{\rm e} = -1$), $\sigma_{\beta \rm  n}$ is the momentum transfer cross section for collisions involving neutrals,  and $\ln \Lambda_{\beta\beta'}$ is Coulomb's logarithm given by \citep[e.g.,][]{spitzer1962,vranjes2013}
\begin{equation}
\ln\Lambda_{\beta\beta'} = \ln \left[ \frac{12\pi \epsilon_0^{3/2}  k_{\rm B}^{3/2} \left( T_\beta + T_{\beta'} \right)}{\left| Z_\beta Z_{\beta'} \right| e^3}  \left( \frac{T_\beta T_{\beta'}}{Z_\beta^2 n_{\beta} T_{\beta'} + Z_{\beta'}^2 n_{\beta'} T_\beta}\right)^{1/2} \right].
\end{equation}
The expressions of the friction coefficients given above are also valid for self-collisions, i.e., collisions between particles of the same species. If for simplicity we assume the same temperature for ions, electrons, and neutrals, i.e., $T_{\rm i} = T_{\rm e} = T_{\rm n}$, and take into account that  the number density of ions and electrons is the same in a hydrogen plasma to satisfy quasi-neutrality, i.e., $n_{\rm i} = n_{\rm e}$, we may drop the subscripts from the temperatures and from Coulomb's logarithm. In addition, we can substitute the numerical values of the various constant parameters into the expression of  Coulomb's logarithm, so that it reduces to \citep[e.g.,][]{priest1984}
\begin{equation}
\ln\Lambda \approx 30.5 - 1.15\log_{10} n_{\rm e} + 3.45 \log_{10} T.
\end{equation}

We are interested in studying the damping of Alfv\'en waves in the chromosphere. Specifically, we are interested in the role of ion-neutral collisions. It is known from previous works \citep[e.g.,][]{zaqarashvili2011,soler2013} that the damping is most efficient when the wave frequency and the ion-neutral collision frequency are of the same order. We are therefore interested in frequencies near the ion-neutral collision frequency. It is important to known whether the fluid theory used here is applicable or breaks down for those wave frequencies. The fluid theory implicitly assumes that self-collisions are frequent enough for fluid behavior to be established in the three fluids separately. This imposes a minimum value for the self-collision frequency to keep the velocity distribution close to a Maxwell–Boltzmann distribution. In other words, the self-collision frequency of a particular species must be higher than the wave frequency and also higher than any of the collision frequencies with the other species. 

The collision frequencies between charged species are essentially determined by the densities and the temperature (see Equation~(\ref{eq:fric})), while the collision frequencies involving neutrals also depend on the neutral collision cross section (see Equation~(\ref{eq:fricneu})). The classical approach to compute the collision cross section is  the so-called model of hard spheres, from here on HS. In the HS model the  particles are considered as solid spheres that interact by means of direct impacts only \citep[e.g.,][]{chapman1970}. The HS cross section is usually computed as $\sigma_{\beta\beta'}=\pi \left( r_\beta + r_{\beta'} \right)^2$, where $r_\beta$ and $r_{\beta'}$ are the radii of particles $\beta$ and $\beta'$, respectively. In the case of ion-neutral and electron-neutral collisions, the radii of both ions and electrons are much smaller than the radius of neutral atoms, so that their HS cross sections are approximately the same, namely $\sigma_{\rm in} \approx \sigma_{\rm en} \approx 10^{-20}$~m$^{2}$. Likewise, the HS cross section for neutral-neutral collisions is $\sigma_{\rm nn} \approx 4\times 10^{-20}$~m$^2$.   

Recently,  \citet{vranjes2013}, from here on VK, presented quantum-mechanical computations of collision cross sections that include several important ingredients missing from the classic HS model. For instance, VK  considered variations of the cross section with temperature, quantum indistinguishability corrections, and charge transfer. The VK cross sections  coincide with the classical ones at  high temperatures, but are different from the HS  cross sections at low temperatures akin to those in the chromosphere. In their paper, VK plot the computed cross section as a function of the energy of the colliding species, which is related to the temperature. For chromospheric temperatures of interest here, we have to consider the results at low energies. The ion-neutral collision cross section from Figure~1 of VK is $\sigma_{\rm in} \approx 10^{-18}$~m$^{2}$, while the electron-neutral collision cross section from their Figure~4 is $\sigma_{\rm en} \approx 3\times 10^{-19}$~m$^{2}$. Concerning neutral-neutral collisions, VK provide in their Figure~3 different cross sections for momentum transfer and viscosity. The cross section for momentum transfer is $\sigma_{\rm nn} \approx 10^{-18}$~m$^{2}$, while the cross section for viscosity is $\sigma_{\rm nn} \approx 3 \times 10^{-19}$~m$^{2}$. In summary, the VK cross sections are between one and two orders of magnitude larger than the classic HS cross sections.

 \begin{figure}
   \centering
  \includegraphics[width=0.95\columnwidth]{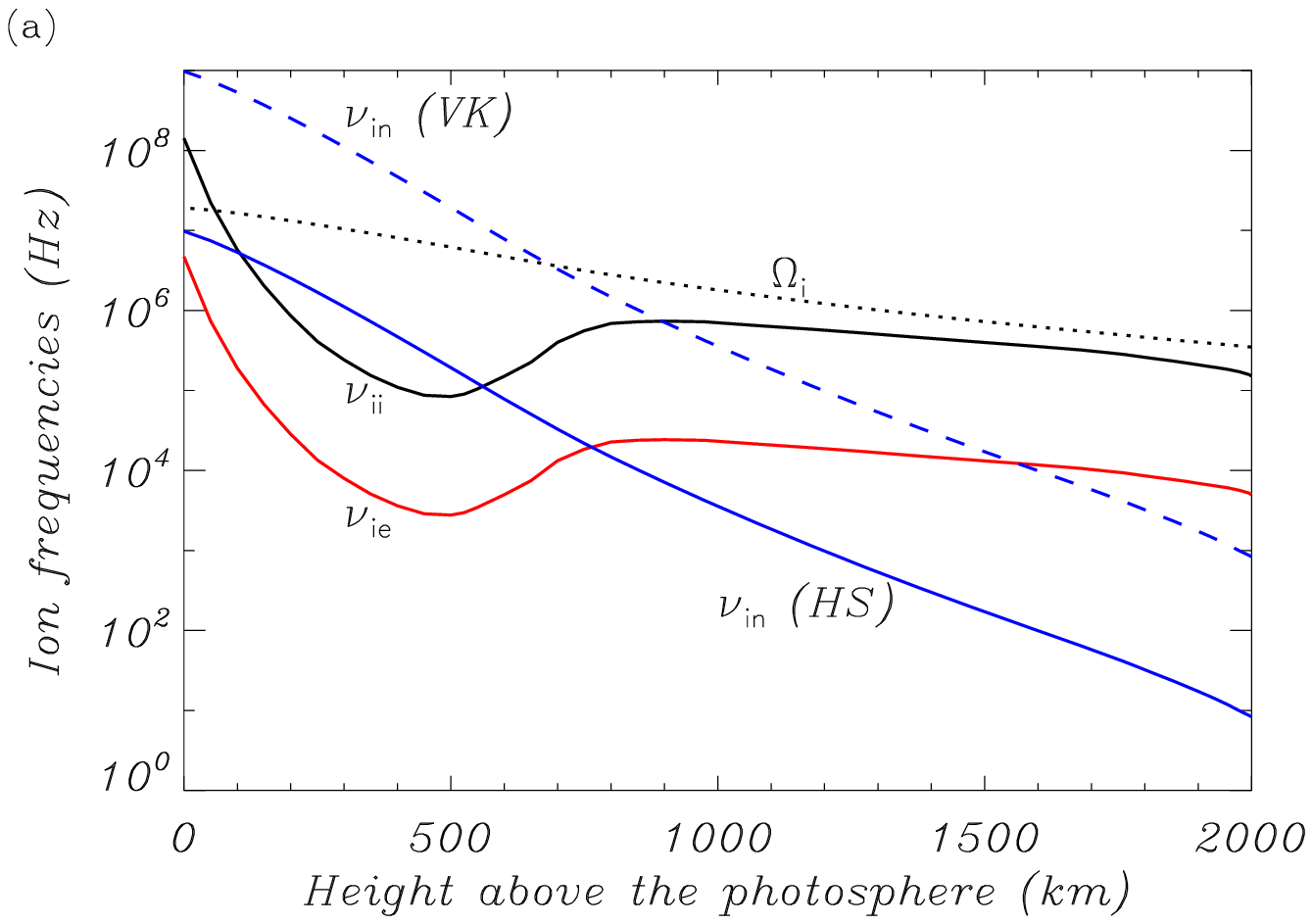}
    \includegraphics[width=0.95\columnwidth]{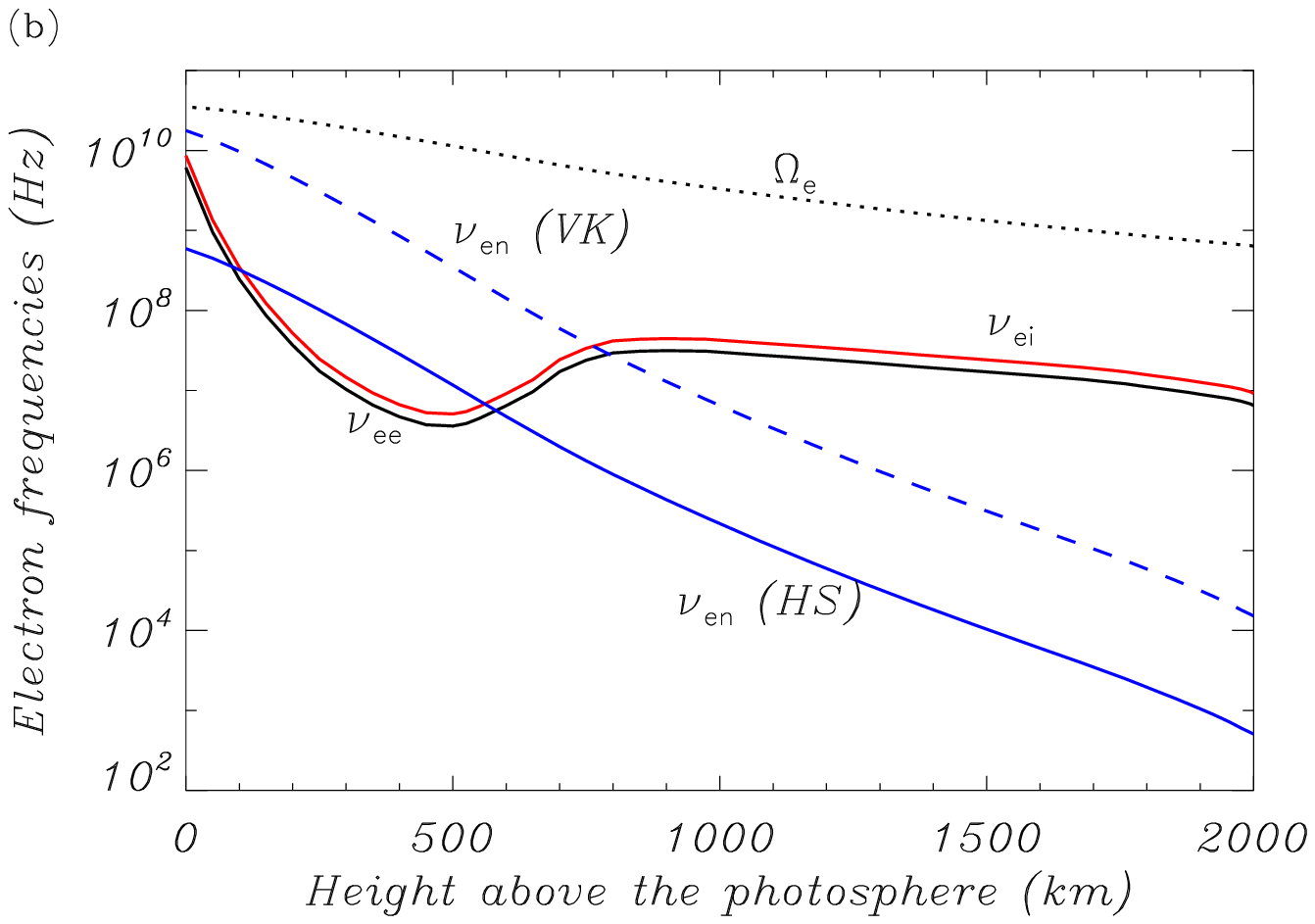}
  \includegraphics[width=0.95\columnwidth]{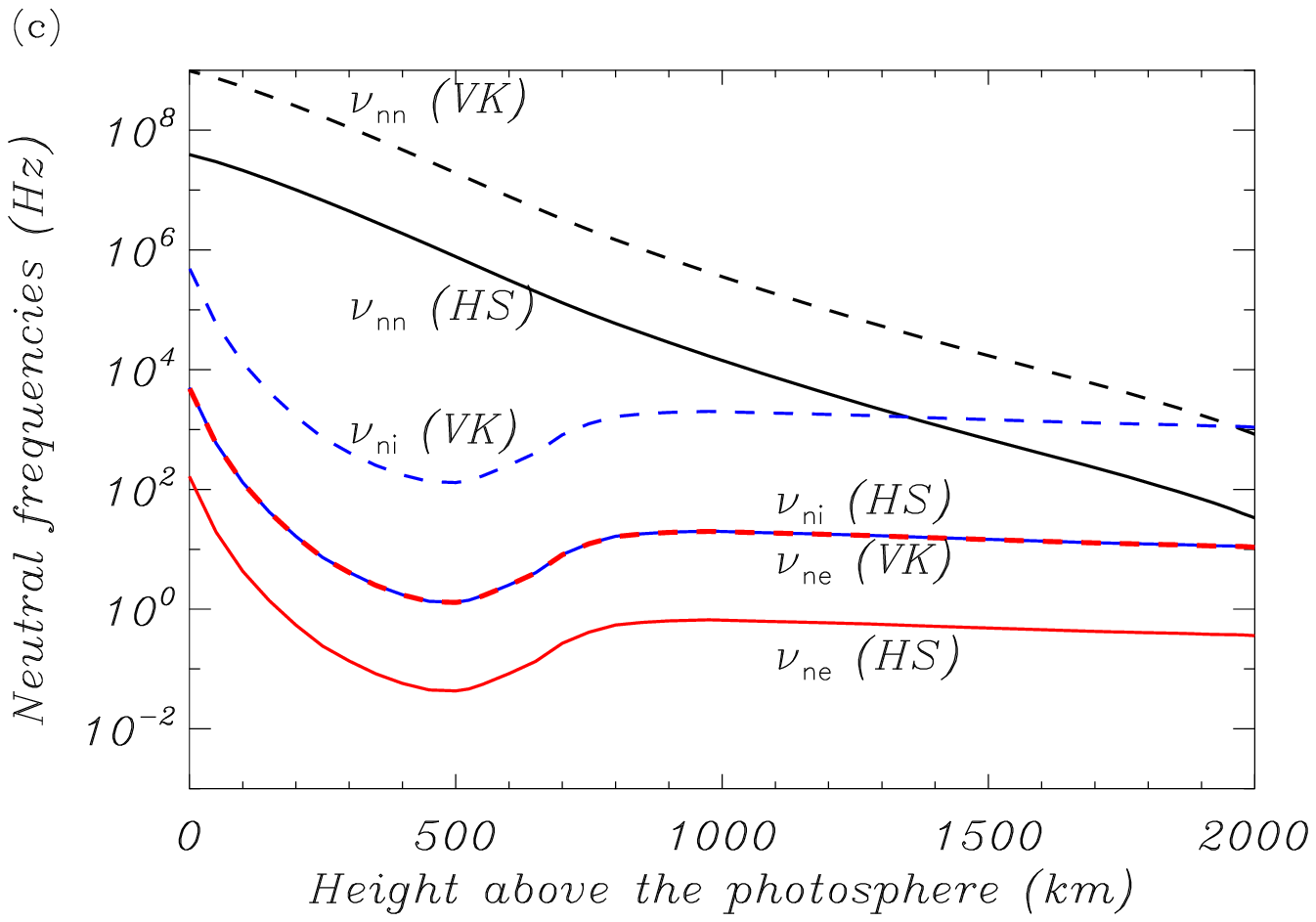}
   \caption{Dependence on height  above the solar photosphere of the collision frequencies for (a) ions, (b) electrons, and (c) neutrals. HS and VK denote collision frequencies with neutrals computed with the hard sphere cross sections and the \citet{vranjes2013} cross sections, respectively. In panels (a) and (b), $\oci$ and $\oce$ denote the ion and electron cyclotron frequencies, respectively.}
              \label{fig:falc2}%
    \end{figure}

The value of the cross section is not only important from the academic point of view but also from the practical point of view. The reason is that the value of the cross section directly affects the collision frequency and, therefore, the cross section is important to determine both the applicability of the fluid theory and the optimal frequency for wave damping. Uncertainties in the collision frequency also cause uncertainties in the various transport coefficients that govern basic collisional phenomena in the plasma \citep[e.g.,][]{martinez2012}. The value of the collision cross section may play an important role in theoretical computations.

Figure~\ref{fig:falc2} shows the dependence on height of the relevant collision frequencies for ions, electrons, and neutrals taking into account both HS and VK cross sections. Concerning ions (Figure~\ref{fig:falc2}(a)), we find that $\nu_{\rm in} \gg \nu_{\rm ii}$ in the low chromosphere.  Collisions with neutrals are too frequent for ions to reach a Maxwell–Boltzmann distribution independently.  In other words, ions collide too frequently with neutrals so that ion-ion collisions do not have enough time to make the ion distribution Maxwellian on their own without the influence of neutrals. Ions are too coupled with neutrals. This means that the condition for ions to be treated as an individual fluid is not satisfied. The multi-fluid theory breaks down  for $h \lesssim 600$~km using the HS cross sections and  for $h \lesssim 900$~km using the VK cross sections. We also see that ions would not be magnetized for $h \lesssim 700$~km in the VK case since $\nu_{\rm in} > \oci$ at those low heights, where $\oci = eB/m_{\rm i}$ is the ion cyclotron frequency \citep[see a discussion on this issue in][]{vranjes2008}. The dependence on height of the ion collision frequencies shown here could be compared to those plotted in VK. However, we note that the magnetic field model used here is different from that of VK. This causes our ion cyclotron frequency to be slightly larger than in VK. We also note that VK used the atmospheric model C of \citet{fontela1993}, while here we use model F. This  leads to a different dependence of the physical parameters (and so the collision frequencies) on height. Concerning magnetization, VK argue that the largest collision frequency of all the possible ones should be used to discuss magnetization. According to VK (see their Figure 7), that collision frequency corresponds to the collision frequency for elastic scattering. However, we note that in Figure 7 of VK the collision frequency for elastic scattering is only slightly larger than that for momentum transfer used here. The use of any of these two frequencies leads essentially to the same results concerning magnetization of ions.

In the case of electrons (Figure~\ref{fig:falc2}(b)), we also see that electrons should not be treated as a separate fluid at low heights due to the very frequent collisions with neutrals, i.e., $\nu_{\rm en} \gg \nu_{\rm ee}$. At large heights, however, it is found that $\nu_{\rm ee} \gg \nu_{\rm en}$ and $\nu_{\rm ee}\sim\nu_{\rm ei}$, so that electrons are strongly coupled to ions (they effectively behave as an ion-electron single fluid) but are weakly affected by neutrals.  In all cases, the electron cyclotron frequency, $\oce = e B/m_{\rm e}$, remains larger than the electron-electron collision frequency. As before, we could compare the electron frequencies computed here to those given in VK. We note again that both the magnetic field dependence on height and the atmospheric model are different in the present paper and in VK. This leads to variations in the physical parameters that may explain the small differences between our Figure~\ref{fig:falc2}(b) and the corresponding graphs of VK.

Finally, the collision frequencies for neutrals (Figure~\ref{fig:falc2}(c)) show that $\nu_{\rm nn}$ is always the largest frequency at all heights, so that treating neutrals as a separate fluid is a valid assumption. This last result remains the same for both HS and VK cross sections. 

The results discussed in the previous paragraphs indicate that for the wave frequencies of interest here, i.e., for wave frequencies near the ion-neutral collision frequency, ions should not be considered as an individual fluid separate from neutrals at low heights in the chromosphere. This was previously noted by \citet{vranjes2013}. The same restriction applies to electrons. The fluid theory is applicable for $h \gtrsim 600$~km using the HS cross sections and for $h \gtrsim 900$~km using the VK cross sections.  Hence, the use of the more accurate VK cross sections results in a more restrictive criterion for the applicability of the fluid theory than that obtained with the classical HS cross sections. The correct study of high-frequency waves in the low chromosphere should be done using hybrid fluid-kinetic models, or even fully kinetic models, since standard fluid theory is not applicable for those high-frequencies. Here we  consider sufficiently large heights in the chromosphere for the fluid theory to be applicable for the wave frequencies under study. Of course, for wave frequencies lower than all the collision frequencies the plasma dynamics can be studied using the single-fluid approximation \citep[e.g.,][]{depontieu2001,khodachenko2004}, which assumes that ions, electrons, and neutrals are strongly coupled. However, here we are interested in higher frequencies beyond the range of applicability of the single-fluid approximation.

\section{Basic equations}
\label{sec:twofluid}

\subsection{Three-fluid model for the upper chromosphere}

 The general multi-fluid equations for a partially ionized plasma can be found in, e.g., \citet{zaqarashvili2011}, \citet{meier2012}, \citet{khomenko2014} among others. Here, we restrict ourselves the linearized version of the equations, that govern the dynamics of small perturbations superimposed on the equilibrium state. This approach is appropriate to study Alfv\'en waves whose velocity amplitudes are much smaller than the Alfv\'en velocity in the plasma. Since Alfv\'en waves  are strictly polarized in the plane perpendicular to the direction of the magnetic field, which is vertical in the present model,  gravity has no effect on the perturbations. In addition, we restrict ourselves to values of the  wavelength that are much shorter than the stratification scale height, and so we perform a local analysis of the perturbations. A limitation of the local analysis is that possible cut-off frequencies and reflection due to the background gravitational stratification are absent \citep[see, e.g.,][]{roberts2006}.  This last issue was recently explored analytically by \citet{zaqarashvili2013} and using numerical simulations by \citet{tu2013}. It is beyond the purpose of the present investigation to take into account the effects of gravitational cut-offs and wave reflection. The reason is that gravitational cut-off may affect waves with longer wavelengths than those studied here.

The linearized momentum equations of ions, electrons, and neutrals are
\begin{eqnarray}
\rhoi \frac{\pd \vi}{\pd t} & = & - \nabla  p_{\rm i}'  - \nabla \cdot \hat{\pi}_{\rm i}+ e n_{\rm i} \left( {\bf E}' + \vi \times {\bf B} \right)  \nonumber \\ &&  - \ain \left( \vi - \vn \right) - \aie \left( \vi - \ve \right) , \label{eq:momion} \\
\rhoe \frac{\pd \ve}{\pd t} & = & - \nabla p_{\rm e}'  - \nabla \cdot \hat{\pi}_{\rm e}  - e n_{\rm e} \left( {\bf E}' + \ve \times {\bf B} \right)  \nonumber \\ &&  - \aen \left( \ve - \vn \right) - \aie \left( \ve - \vi \right), \label{eq:momelec} \\
\rhon \frac{\pd \vn}{\pd t} & = &  -\nabla p_{\rm n}' - \nabla \cdot \hat{\pi}_{\rm n} - \ain \left( \vn - \vi \right) - \aen \left( \vn - \ve \right) , \label{eq:momneu} 
\end{eqnarray}
where ${\bf v}_\beta$, $p_\beta'$, $\hat{\pi}_\beta$, and $\rho_\beta = m_\beta n_\beta$ are the velocity perturbation, scalar pressure perturbation, viscosity tensor, and mass density of species $\beta$, and $\bf E'$ is the electric field perturbation.  Since in the present paper we study incompressible Alfv\'en waves,  Equations~(\ref{eq:momion})--(\ref{eq:momneu}) are complemented with the incompressibility conditions for the three fluids, namely
\begin{equation}
\nabla \cdot \vi = \nabla \cdot \ve = \nabla \cdot \vn = 0. \label{eq:incomp}
\end{equation}

For Alfv\'en waves the $z$-components of the velocity perturbations are zero, i.e., $v_{{\rm i},z}=v_{{\rm e},z}=v_{{\rm n},z}=0$. This means that it is enough take the $x$- and $y$-components of Equations~(\ref{eq:momion})--(\ref{eq:momneu}). Although we include scalar pressure terms in the momentum equations for completeness, we note that these forces do not act on incompressible Alfv\'en waves and may be dropped. The terms with $\ain$, $\aie$ and $\aen$  in Equations~(\ref{eq:momion})--(\ref{eq:momneu}) account for the transfer of momentum between species due to particle collisions. Concerning viscosity, we take into account that the viscosity of a partially ionized plasma is determined essentially by  ions and  neutrals by virtue of their larger masses \citep[see, e.g.,][]{brag, meier2012}. Therefore, the electron viscosity is neglected  compared to the ion and neutral viscosities.  The ion viscosity tensor is a complicated expression usually described as the sum of five  components accounting for compressive viscosity, shear viscosity, and gyroviscosity \citep[see the full expression in][]{brag}. Keeping the spatial variations (derivatives)  of the velocity along the direction of the magnetic field only, as appropriate for Alfv\'en waves, the $x$- and $y$-components of the divergence of the ion viscosity tensor  reduce to
\begin{eqnarray}
\left( \nabla \cdot \hat{\pi}_{\rm i} \right)_x &=& - \rhoi \left( \xi_{\rm i,\perp} \frac{\partial^2 v_{{\rm i},x}}{\partial z^2} +  \xi_{\rm i,g}  \frac{\partial^2 v_{{\rm i},y}}{\partial z^2} \right), \\
\left( \nabla \cdot \hat{\pi}_{\rm i} \right)_y &=& - \rhoi \left( \xi_{\rm i,\perp} \frac{\partial^2 v_{{\rm i},y}}{\partial z^2} -  \xi_{\rm i,g}  \frac{\partial^2 v_{{\rm i},x}}{\partial z^2} \right), 
\end{eqnarray}
with $\xi_{\rm i,\perp}$ and $\xi_{\rm i,g}$ the reduced coefficients of  ion shear viscosity and gyroviscosity, respectively, which are adapted from \citet{brag}, namely
\begin{eqnarray}
\xi_{\rm i,\perp} &=& \frac{6}{5} \frac{ n_{\rm i} k_{\rm B} T \nu_{\rm ii}}{\rhoi \oci^2} = \frac{6}{5} \frac{ c_{\rm T,i}^2 \nu_{\rm ii}}{\oci^2}, \\
\xi_{\rm i,g} &=&  \frac{ n_{\rm i} k_{\rm B} T }{\rhoi \oci} = \frac{c_{\rm T,i}^2}{\oci},
\end{eqnarray}
where  $c_{\rm T,i} = \sqrt{k_{\rm B} T/m_{\rm i}}$ is the ion thermal velocity. We note that the coefficient of ion compressive viscosity \citep[see][]{brag} does not play a role in Alfv\'en waves. The reason for this result is that Alfv\'en waves are incompressible and produce no velocity perturbations along the magnetic field direction. In the neutral fluid the form of the viscosity tensor is simpler because of the absence of the effect of the magnetic field. Thus, the divergence of the neutral viscosity tensor  is
\begin{equation}
\nabla \cdot \hat{\pi}_{\rm n} = - \rhon \xi_{\rm n} \nabla^2 \vn, \label{eq:viscosity}
\end{equation} 	
with $\xi_{\rm n}$ the isotropic coefficient of neutral viscosity given by
\begin{equation}
 \xi_{\rm n} = \frac{2 n_{\rm n} k_{\rm B} T}{\rhon \nu_{\rm nn}} =  \frac{2 c_{\rm T,n}^2}{\nu_{\rm nn}},
\end{equation}
where  $c_{\rm T,n} = \sqrt{k_{\rm B} T/m_{\rm n}}$ is the neutral thermal velocity. Since $m_{\rm n} \approx m_{\rm i}$ for hydrogen, $c_{\rm T,i} \approx c_{\rm T,n}$ and so we can drop the subscript from the thermal velocity.

\subsection{Reduction of the main equations}

From here on we specialize in Alfv\'enic perturbations, and so we drop  pressure perturbations. It is frequent in the literature to exclude  the electric field perturbation, $\bf E'$, from the equations and to work with  the magnetic field perturbation, $\bf B'$, instead. In that case, the induction equation governing the evolution of the magnetic field perturbation has to be obtained  \citep[see, e.g.,][]{zaqarashvili2011}. It is also useful to manipulate the equations so that the electron velocity does not explicitly appear. To do so,  we define the current density perturbation as
\begin{equation}
{\bf j} =  e \left( n_{\rm i} \vi - n_{\rm e} \ve \right),
\end{equation}
which is related to the magnetic field perturbation as
\begin{equation}
{\bf j} = \frac{1}{\mu} \nabla \times {\bf B}',
\end{equation}
where $\mu$ is the magnetic permeability. These relations are used to express the electron velocity, $\ve$,  as
\begin{equation}
\ve = \vi - \frac{1}{\mu e n_{\rm e}}  \nabla \times {\bf B}', \label{eq:ve}
\end{equation}
where we used the quasi-neutrality condition $n_{\rm i} = n_{\rm e}$. This equation governs the electron dynamics and replaces their full momentum equation (Equation~(\ref{eq:momelec})). In turn, Equations~(\ref{eq:momion})--(\ref{eq:momneu}) and (\ref{eq:ve}) can be conveniently combined to eliminate $\bf E'$ and $\ve$ from the ion and neutral momentum equations, namely
\begin{eqnarray}
\rhoi \frac{\partial \vi}{\partial t} &=& - \nabla \cdot \hat{\pi}_{\rm i}  + \frac{1}{\mu} \left( \nabla \times {\bf B}' \right) \times {\bf B} - \left( \ain + \aen \right) \left( \vi - \vn \right)  \nonumber \\
&&+ \frac{m_{\rm e}}{\mu e}\left( \frac{\aen}{m_{\rm e} n_{\rm e}} + \frac{\partial}{\partial t} \right)\nabla \times {\bf B}', \label{eq:momionfin} \\
\rhon \frac{\partial \vn}{\partial t} &=& - \nabla \cdot \hat{\pi}_{\rm n} - \left( \ain + \aen \right) \left( \vn - \vi \right) -\frac{\aen}{\mu e n_{\rm e}} \nabla \times {\bf B}', \label{eq:momneufin}
\end{eqnarray}
where we approximated $\rhoi+\rhoe \approx \rhoi$.

Next, the expression for the electric field perturbation, $\bf E'$, is obtained from Equation~(\ref{eq:momelec}), namely
\begin{eqnarray}
{\bf E'} &=& -\vi \times {\bf B} + \frac{\aie+\aen}{\mu e^2 n_{\rm e}^2} \nabla \times {\bf B}' + \frac{1}{\mu e n_{\rm e}} \left(  \nabla \times {\bf B}'  \right) \times {\bf B} \nonumber \\
&& - \frac{m_{\rm e}}{e}\frac{\partial}{\partial t} \left( \vi - \frac{1}{\mu e n_{\rm e}} \nabla \times {\bf B}' \right) + \frac{\aen}{e n_{\rm e}} \left(  \vn - \vi \right).
\end{eqnarray}
Finally,  we use the Maxwell Equation,
\begin{equation}
\frac{\partial {\bf B}'}{\partial t} = - \nabla \times {\bf E}',
\end{equation}
 and arrive at the  equation governing the evolution of the magnetic field perturbation, namely
\begin{eqnarray}
\frac{\pd {\bf B}'}{\pd t} & = & \nabla \times \left(  \vi \times {\bf B} \right)  - \left( \eta_0 + \eta_1 \frac{\pd }{\pd t}  \right)  \nabla \times \nabla \times {\bf B}' + \frac{m_{\rm e}}{e} \frac{\pd }{\pd t} \nabla \times \vi \nonumber \\
&& - \eta_{\rm H}\nabla \times \left[ \left( \nabla \times {\bf B}' \right) \times {\bf B} \right]  -\eta_2 \nabla \times \left( \vn-\vi \right). \label{eq:indu2f}
\end{eqnarray}
Equation~(\ref{eq:indu2f}) is the linearized induction equation in a pressureless three-fluid partially ionized plasma. The terms on the right-hand side of Equation~(\ref{eq:indu2f}) are the  inductive term,  Ohm's magnetic diffusion, a term due to electron inertia, Hall's current, and a term that mostly accounts for the effect of electron-neutral collisions, respectively. We define the various coefficients in Equation~(\ref{eq:indu2f}) as follows
\begin{eqnarray}
\eta_0 &=& \frac{\aie+\aen}{\mu e^2 n_{\rm e}^2} , \\
\eta_1 &=& \frac{m_{\rm e}}{\mu e^2 n_{\rm e}},\\
 \eta_{\rm H} &=& \frac{1}{\mu e n_{\rm e}}, \\
 \eta_2 &=& \frac{\aen}{e n_{\rm e}}.
\end{eqnarray}
Ohm's term and the term due to electron-neutral collisions represent diffusion of the magnetic field due to collisions between particles.  Hall's current and the electron inertia term are not a dissipative terms. The so-called Hall effect arises in a plasma when electrons are able to drift with the magnetic field but the much heavier ions are not completely frozen to the magnetic field. As a result, the current density vector has a component normal to the electric field vector \citep[see, e.g.,][]{priest1984}.  Hall's effect may be enhanced by electron-neutral collisions, which tend to further decouple ions and electrons \citep{pandey2008}.

\subsection{Dispersion relation and the quality factor}

To obtain the dispersion relation of Alfv\'en waves, we take into account that any disturbance in the plasma can be expressed as a superposition of its Fourier modes and, for linear waves, it is enough studying the behavior of the Fourier modes separately. Thus, we  put all perturbations proportional to $\exp\left( i k z - i\omega t \right)$, where  $k$ is the wavenumber along the magnetic field direction, and $\omega$ is the angular frequency. The wavelength, $\lambda$, and the wave frequency, $f$, are related to these two parameters as $\lambda = 2\pi/k$ and $f = \omega/2\pi$.

Equations~(\ref{eq:momionfin}), (\ref{eq:momneufin}) and (\ref{eq:indu2f}) define an eigenvalue problem where the $x$- and $y$-components of $\vi$, $\vn$, and $\bf B'$ form the eigenvector, and either $\omega$ or $k$ is the eigenvalue. The $z$-components of $\vi$, $\vn$, and $\bf B'$ are zero for Alfv\'en waves. The eigenvalue is the solution of the  dispersion equation that relates the  frequency to the wavenumber. The dispersion relation is here expressed as a 6x6 determinant, namely
\begin{equation}
 \left| 
\begin{array}{cccccc}
\Gamma_{\rm i} & -k^2 \xi_{\rm i,g} & \frac{\ain+\aen}{\rhoi} & 0 & \frac{ikB}{\mu \rhoi} & - \Sigma \\
k^2 \xi_{\rm i,g} & \Gamma_{\rm i} & 0 & \frac{\ain+\aen}{\rhoi} & \Sigma & \frac{ikB}{\mu \rhoi} \\
\frac{\ain+\aen}{\rhon} & 0 & \Gamma_{\rm n} & 0 & 0 & \frac{ik\aen}{\mu \rhon e n_{\rm e}} \\
0 & \frac{\ain+\aen}{\rhon} & 0 & \Gamma_{\rm n} &   -\frac{ik\aen}{\mu \rhon e n_{\rm e}} & 0 \\
i k B & -\Psi & 0 & i k \eta_2 & \Gamma_{\rm b} & -k^2 B \eta_{\rm H} \\
\Psi & i k B & - i k \eta_2 & 0 & k^2 B \eta_{\rm H} & \Gamma_{\rm b}
\end{array}
 \right| = 0, \label{eq:reldisper}
\end{equation}
with
\begin{eqnarray}
\Gamma_{\rm i} &=& i \omega - k^2 \xi_{\rm i,\perp}  - \frac{\ain + \aen}{\rhoi}, \\
\Gamma_{\rm n} &=& i \omega - k^2 \xi_{\rm n}  - \frac{\ain + \aen}{\rhon}, \\
\Gamma_{\rm b} &=& i \omega - k^2 \left( \eta_0 - i\omega \eta_1 \right), \\
\Sigma &=& i k \frac{m_{\rm e}}{\mu \rhoi e} \left( \frac{\aen}{m_{\rm e}n_{\rm e}} - i\omega \right), \\
\Psi &=& k \left( \omega \frac{m_{\rm e}}{e} + i\eta_2 \right).
\end{eqnarray}

We consider the paradigmatic case that  the effects of partial ionization, viscosity, Ohm's diffusion, and Hall's current  are all dropped from the dispersion relation. In that case, the dispersion relation reduces to a 2x2 determinant, namely 
\begin{equation}
\left| \begin{array}{cc}
i \omega &  \frac{ikB}{\mu \rhoi} \\
i k B & i \omega
\end{array}  \right| = 0,
\end{equation}
whose solutions  are the well-known ideal linearly polarized Alfv\'en waves in a fully ionized plasma, namely
\begin{equation}
\omega = \pm k \vai,
\end{equation}
where $\vai = B/\sqrt{\mu\rhoi}$ is the ion Alfv\'en velocity, and the $+$ and $-$ signs stand for forward and backward propagating waves, respectively.   The ideal Alfv\'en waves are modified when non-ideal mechanisms are taken into account. 

Alfv\'en waves efficiently  transport energy in magnetized plasmas. Due to the presence of dissipation mechanisms, wave energy can be deposited in the plasma. For instance,  ion shear viscosity,  neutral isotropic viscosity,  collisions, and Ohm's diffusion are dissipation mechanisms that produce the damping of the waves. Hall's current and ion gyroviscosity, however, do not produce dissipation but dispersion. Hall's current and gyroviscosity are able to break the symmetry between the two possible circular polarizations of the electric field, namely left and right circular polarizations \citep[see, e.g.,][]{zhelyazkov1996,cramer2001}, which increases the number of different waves that can propagate in the plasma. Formally, a linearly polarized Alfv\'en wave can be written as a sum of a left circularly polarized wave and a right circularly polarized wave, with the implicit assumption that both waves have the same frequency \citep[see, e.g.,][]{pecseli2013}. When the two circular polarizations have different frequencies  a pure linearly polarized Alfv\'en wave cannot be formed, and so the two waves with opposite circular polarization appear as different solutions of the dispersion relation.

 The dispersion relation (Equation~(\ref{eq:reldisper})) applies to both the spatial and the temporal regimes, depending on whether $\omega$ or $k$ is prescribed. On the one hand, the solutions in the temporal regime can be related to the impulsively driven or initial-value problem, in which the waves are excited by an impulsive driver of short duration that generate local perturbations in the plasma with a certain spatial extent. In the temporal regime the wavenumber, $k$, is a prescribed real quantity and the  angular frequency, $\omega$, is obtained from the dispersion relation. Due to the presence of dissipation mechanisms the waves are damped in time. Temporal damping is mathematically represented by the fact that  $\omega$ is complex. The real part of $\omega$ is related to the oscillatory behavior of the wave, i.e., the period.  In turn, the imaginary part of $\omega$ corresponds to the temporal damping rate. On the other hand, the solutions in the spatial regime can be linked to the periodically driven or boundary-value problem, in which the waves are excited by a periodic driver with a certain frequency.  In the spatial regime  $\omega$ is a prescribed real quantity and $k$ is the complex solution of the dispersion relation. In this case, the real part of $k$ is related to the actual wavelength and the imaginary part of $k$ corresponds to the spatial damping length.

The quality factor, $Q$, is a dimensionless parameter that characterizes how efficiently damped a wave is. In the case of temporal damping for a prescribed $k$,  the standard definition of the quality factor is
\begin{equation}
Q \equiv \frac{1}{2}\left| \frac{{\rm Re}(\omega)}{{\rm Im}(\omega)} \right|. \label{eq:quality}
\end{equation}
The quality factor compares the frequency at which a wave oscillates to the rate at which it damps. Equivalently, in the case of spatial damping for a prescribed $\omega$, the quality factor is
\begin{equation}
Q\equiv \frac{1}{2}\left| \frac{{\rm Re}(k)}{{\rm Im}(k)} \right|, \label{eq:qualityspatial}
\end{equation}
so that $Q$ compares the wavelength with the damping length. The behavior of the waves depends on the value of $Q$. When $Q>1/2$  perturbations are underdamped, meaning that the perturbations oscillate while their amplitude  decrease in time or space. Most of the energy of the perturbations can propagate away  as  weakly damped  Alfv\'en waves. The larger $Q$, the weaker the damping and so the farther the wave can propagate from the excitation location until all the wave energy is dissipated in the plasma. If $Q \to \infty$, the wave is undamped and no energy dissipation takes place. Conversely, when $Q < 1/2$ the dissipation is very strong and the perturbation is overdamped. Most of the energy stored in the perturbation is dissipated {\em in situ} instead of being transported away in the form of a propagating Alfv\'en wave. The case $Q=1/2$ that separates the two regimes is often called critical damping. The most extreme situation, however, takes place when $Q=0$. In such a case, the wave has a so-called cut-off. In a cut-off scenario, perturbations are evanescent in time or space instead of oscillatory and therefore Alfv\'en waves cannot propagate at all. The remainder of this paper is devoted to investigating the quality factor of Alfv\'en waves in the solar chromosphere.

 \section{Alfv\'en wave damping in the upper chromosphere}

\label{sec:damping}

\subsection{Impulsive driver}

\begin{figure*}
   \centering
  \includegraphics[width=.95\columnwidth]{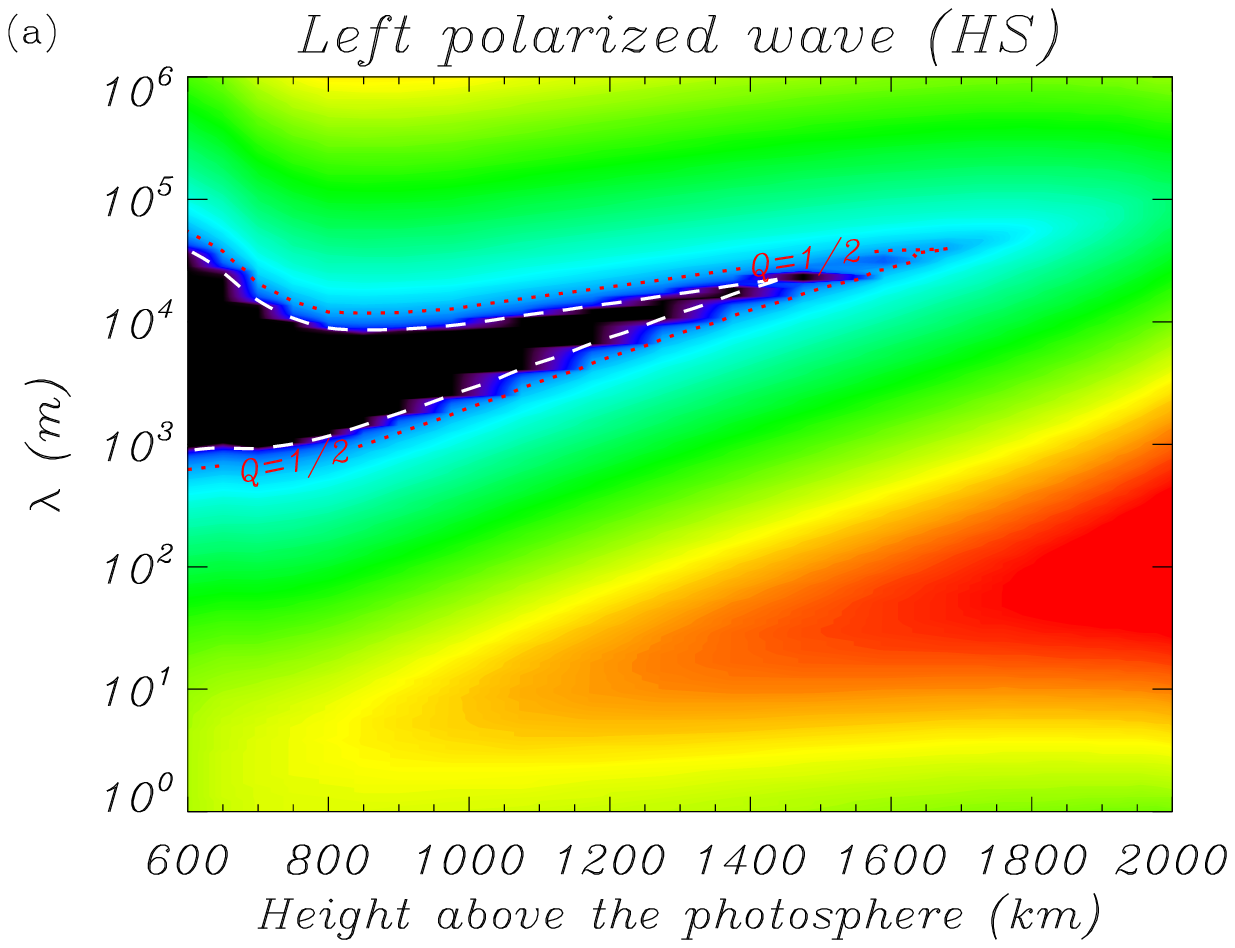}
    \includegraphics[width=.95\columnwidth]{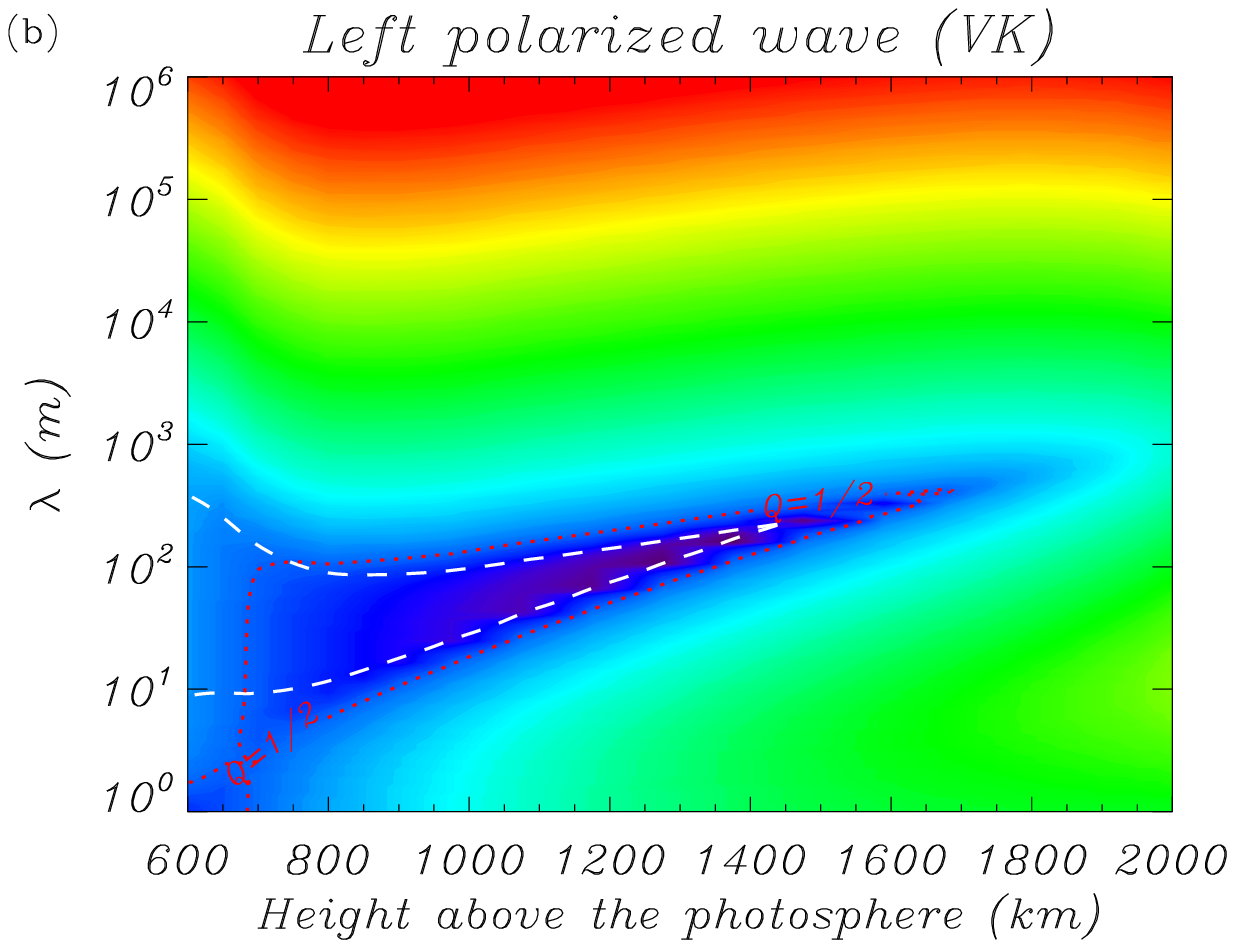}
 \includegraphics[width=.95\columnwidth]{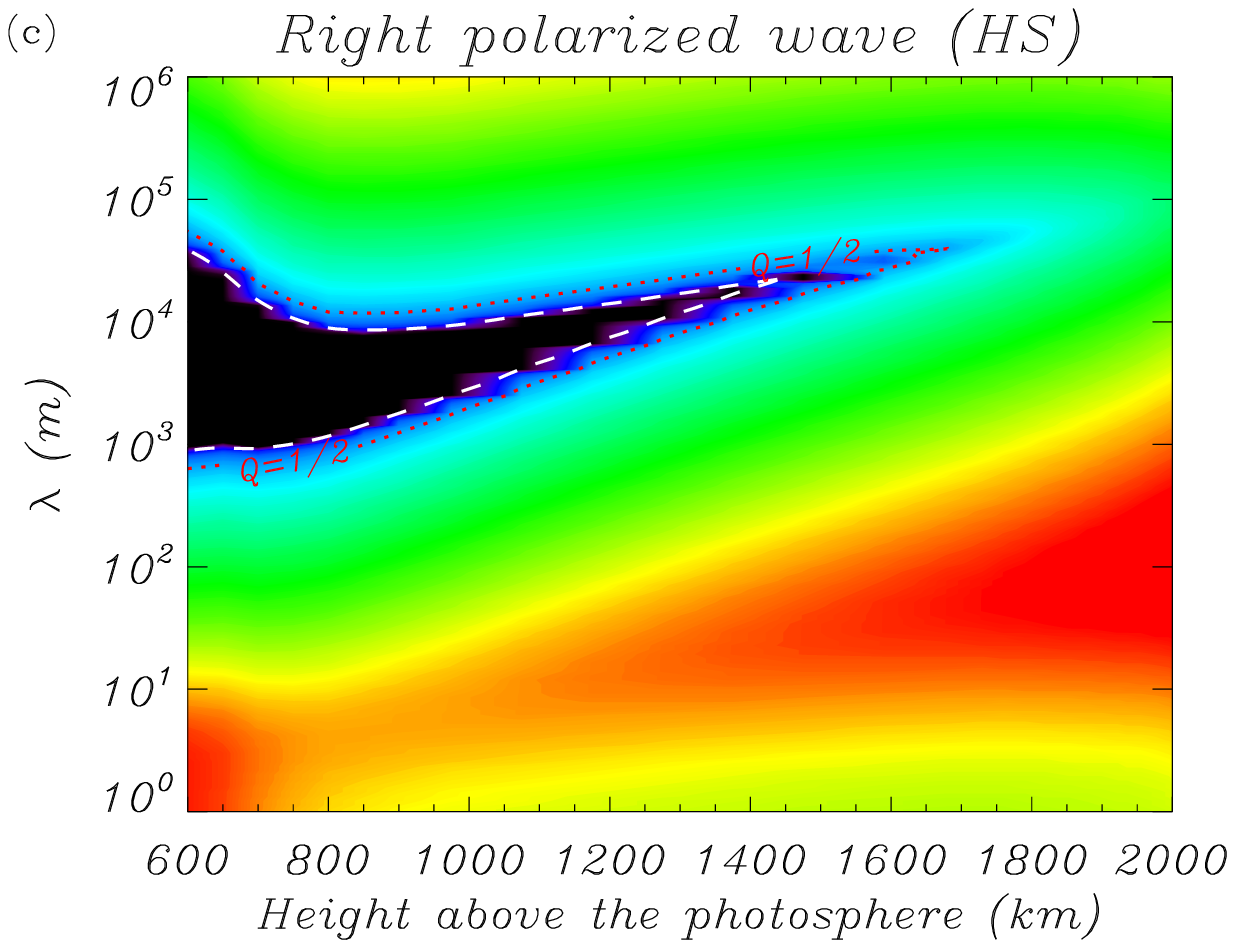}
    \includegraphics[width=.95\columnwidth]{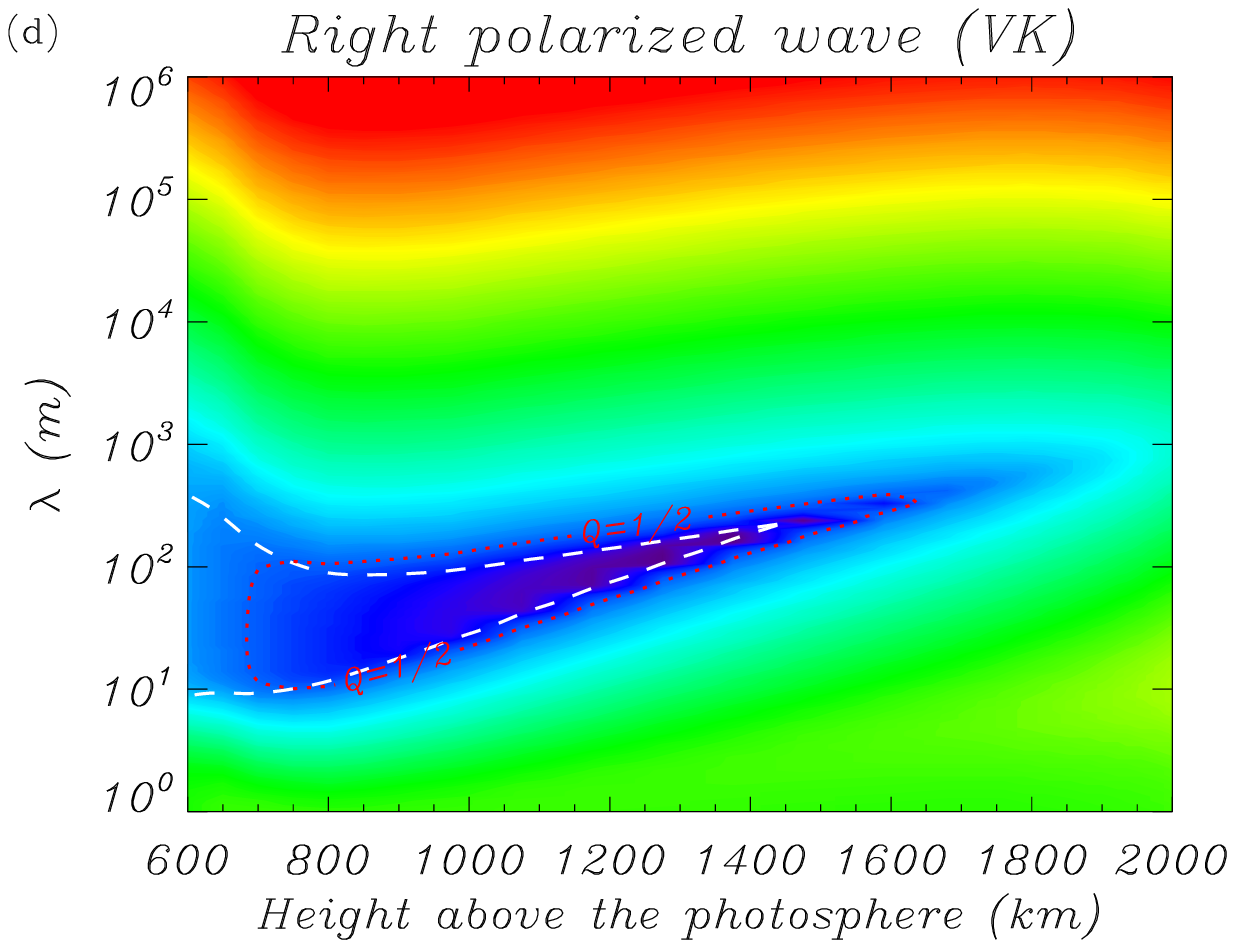}
  \includegraphics[width=1.45\columnwidth]{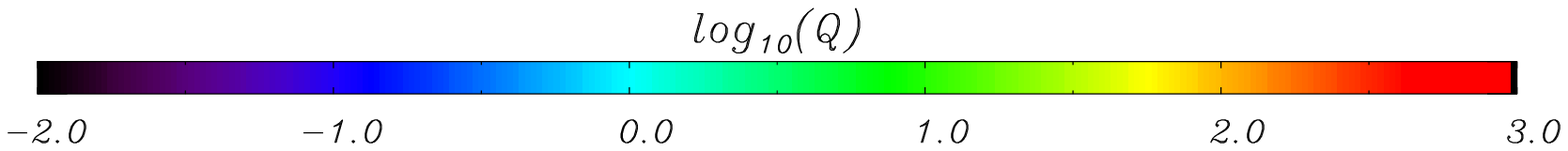}
   \caption{Contour plots of the quality factor $Q$ (in logarithmic scale) of impulsively driven Alfv\'en waves in the upper solar chromosphere as function of the  wavelength $\lambda$ (vertical axis in logarithmic scale) and height above the photosphere (horizontal axis). Top and botton panels correspond to the left and right polarized waves, respectively.  Left and right panels are the results obtained using the hard-sphere  cross sections (HS) and the \citet{vranjes2013} cross sections (VK), respectively. The red dotted lines enclose zones where Alfv\'en waves are overdamped, i.e., $Q < 1/2$. The white dashed lines correspond to the approximate critical wavelengths given in Equations~(\ref{eq:l1}) and (\ref{eq:l2}). We note that the VK results are only strictly valid for heights larger than 900~km.}
              \label{fig:temporal1}%
    \end{figure*}

We start by investigating the case of Alfv\'en waves excited by an impulsive driver. Hence, the temporal damping of the waves for a fixed wavelength is explored. The dispersion relation is solved to obtain the complex wave frequency, and so the quality factor, as a function of height. The dispersion relation has six solutions for the frequency. Four solutions correspond to forward and backward, left and right circularly polarized Alfv\'en waves, while the two remaining solution are  vortex modes \citep[see][]{zaqarashvili2011,soler2013}, also called forced neutral oscillations  \citep{vranjes2014}. Vortex modes are associated with vorticity perturbations in the neutral fluid and do not represent Alfv\'en waves. Hence, the solutions corresponding to vortex modes are discarded from the present study. We focus on studying the quality factor of the Alfv\'en waves.

Figure~\ref{fig:temporal1} shows contour plots of the quality factor (in logarithmic scale) as a function of height, $h$, above the photosphere for  wavelengths, $\lambda=2\pi/k$,  in between 1~m and $10^6$~m. We note that the pressure scale height in the chromosphere is around 300 km, hence wavelengths above $10^5$~m are probably stretching the assumptions behind the local analysis. The quality factor does not show a strong dependence on the wave circular polarization, and both left and right polarized solutions have similar values of $Q$. However, the quality factor is found to be strongly dependent on the set of cross sections used in the computations. Figure~\ref{fig:temporal1} displays some regions where Alfv\'en waves are overdamped, i.e., $Q<1/2$ (see the regions denoted by red dotted lines in the various panels). The position of these regions of overdamping is different depending on the cross sections used. For instance, the region of overdamping takes place at longer wavelengths for the HS cross sections than for the VK cross sections. Hence, the use of the more accurate VK cross sections has here a very strong impact. The presence of  regions of overdamping  has important implications concerning the dissipation of wave energy \citep[see][]{soler2013,soler2013ma}. Perturbations in the plasma whose wavelengths fall within those regions cannot propagate away from the location of the excitation in the form of traveling Alfv\'en waves. Instead, all the energy stored in the perturbations is necessarily dissipated {\em in situ}. This may translate in a strong plasma heating in those locations \citep[see also][]{song2011}.                                                   

The regions of overdamping shown in Figure~\ref{fig:temporal1} can be directly related to the cut-off interval of wavelengths explored by \citet{soler2013}. These authors  found that there is an interval of wavelengths,  namely $\lambda_1 < \lambda < \lambda_2$, for which the ion-neutral friction force becomes dominant over the restoring magnetic tension force. As a consequence, magnetic field perturbations decay in a time scale much shorter than the wave period, which result in the suppression of the Alfv\'en wave magnetic field oscillations. The existence of a critical interval of  wavelengths for Alfv\'en waves in a partially ionized plasma was first reported by \citet{kulsrud1969} and was subsequently investigated by, e.g., \citet{mouschovias1987}, \citet{kamaya1998}, \citet{soler2013,soler2013ma,soler2013tube}, among others.  The overall physical picture taking into account the dynamics of ions, electrons, and neutrals can be summarized as follows. Ions, electrons, and neutrals move as a single fluid when $\lambda > \lambda_2$. When $\lambda_1 < \lambda < \lambda_2$, the strength of friction becomes larger than that of magnetic tension and the fluids cannot oscillate as a whole any more, i.e., magnetic tension is not strong enough to move the whole plasma. When $\lambda < \lambda_1 $ neutrals decouple from ions and electrons, i.e., magnetic tension is again able to put ions and electrons into motion, but neutrals remain  at rest. Thus, when $\lambda < \lambda_1 $ Alfv\'en waves only produce perturbations in the ion and electron fluids, whereas neutrals are not perturbed. The critical interval of wavelengths represents the transition from the regime where ions, electrons, and neutrals support Alfv\'en waves to the regime where  Alfv\'en waves perturb  ions and electrons  only \citep[see an extended discussion in][]{kamaya1998}.

 As explained in \citet{soler2013}, the presence or absence of the critical interval of wavelengths depends upon the value of the ionization fraction, $\chi = \rhon/\rhoi$. The condition to be satisfied is that $\chi > 8$, which takes place for $h\lesssim$~1,500~km in the present chromospheric model. Analytic expressions of $\lambda_1$ and $\lambda_2$ were obtained by \citet{soler2013}, namely
\begin{equation}
\lambda_{1,2} = 2\pi\frac{\chi \vai}{\nuin} \sqrt{ \frac{\chi^2 + 20\chi - 8}{8 \left(1 + \chi \right)^3} \pm \frac{\chi^{1/2}\left(\chi - 8\right)^{3/2}}{8 \left(1 + \chi \right)^3}}, \label{eq:k12}
\end{equation}
where the $-$ and the $+$ signs stand for $\lambda_1$ and $\lambda_2$, respectively. Simplified expressions for these wavelengths can be obtained assuming weak ionization, i.e., $\chi \gg 1$, so that the approximations are
\begin{eqnarray}
 \lambda_1 &\approx & 2\pi\frac{\vai}{\nuin}\frac{2\chi}{\chi + 1} \approx 4\pi\frac{\vai}{\nuin} = \frac{4\pi}{\nuin} \frac{B}{\sqrt{\mu \rhoi}},  \label{eq:l1} \\
  \lambda_2 &\approx & 2\pi\frac{\vai}{\nuin}\frac{\chi}{2\sqrt{\chi+1}} \approx \pi\frac{\vai\sqrt{\chi}}{\nuin} = \frac{\pi}{\nuni}\frac{B}{\sqrt{\mu \rhon}}. \label{eq:l2} 
\end{eqnarray}
These approximate wavelengths agree reasonably well with the boundaries of the regions of overdamping in Figure~\ref{fig:temporal1}. The  differences may be attributed to the additional physical processes included here and that were absent from the analysis of \citet{soler2013}. The ratio of the two critical wavelengths is
\begin{equation}
\frac{ \lambda_2}{ \lambda_1} \approx \frac{1}{4}\sqrt{\frac{\rhon}{\rhoi}}. \label{eq:ratiolam}
\end{equation}
Importantly, $\lambda_2 / \lambda_1$ is independent of the ion-neutral collision cross section and is only  a function of the ratio of neutral to ion densities. Equation~(\ref{eq:ratiolam}) could be used to indirectly estimate the plasma ionization degree if the values of $\lambda_1$ and $\lambda_2$ were inferred from observations.

The effects of Hall's current, electron-neutral collisions, electron inertia, magnetic difussion, and viscosity were absent from the work of \citet{soler2013}. The additional physical effects considered here provide a more realistic representation of the chromospheric plasma  and have an  impact on the behavior of the waves. The presence of Hall's current  and electron inertia cause the strict frequency cut-offs obtained by \citet{soler2013}  to be replaced by zones where Alfv\'en waves are  overdamped.  This result was also discussed in the previous work by  \citet{zaqarashvili2012}. In simple physical terms, the effect of Hall's current and electron inertia for removing the strict frequency cut-off can be understood as follows. In the absence of Hall's current and electron inertia, electrons can be considered as tightly coupled to ions, in the sense that electrons just follow the behavior of ions. Both ions and electrons are frozen into the magnetic field. In this case, ion-neutral collisions can completely suppress the magnetic field perturbations and so cause the cut-off, as discussed in \citet{soler2013}. However, when either Hall's current or electron inertia are included, electrons can have a different dynamics than that of ions. Ions may not be able follow the magnetic field fluctuations due to the effect of ion-neutral collisions, but it is easier for electrons to remain coupled to magnetic field. Therefore, ion-neutral collisions cannot completely suppress the fluid oscillations because of the distinct behavior of electrons when Hall's current and/or electron inertia are included \citep[see also the discussion in][]{pandey2008}.

 The influence of Hall's current also allows us to understand why, within the regions of overdamping, the damping is stronger for the HS model than for the VK model. Hall’s current helps to reduce the efficiency of  damping, i.e., the influence of Hall’s current increases the quality factor. The importance of Hall’s current grows when the wave frequency increases and approaches the ion cyclotron frequency.  Since the region of strongest damping takes place at higher frequencies (shorter wavelengths) in the VK case than in the HS case, the effect of Hall’s current in the regions of overdamping is more important in the VK case than in the HS case. Thus, the strength of the damping appears to be slightly lower in the VK case than in the HS case when the value of the quality factor within  the regions of overdamping of Figure~\ref{fig:temporal1} is compared.

\subsection{Periodic driver}

\begin{figure*}
   \centering
  \includegraphics[width=.95\columnwidth]{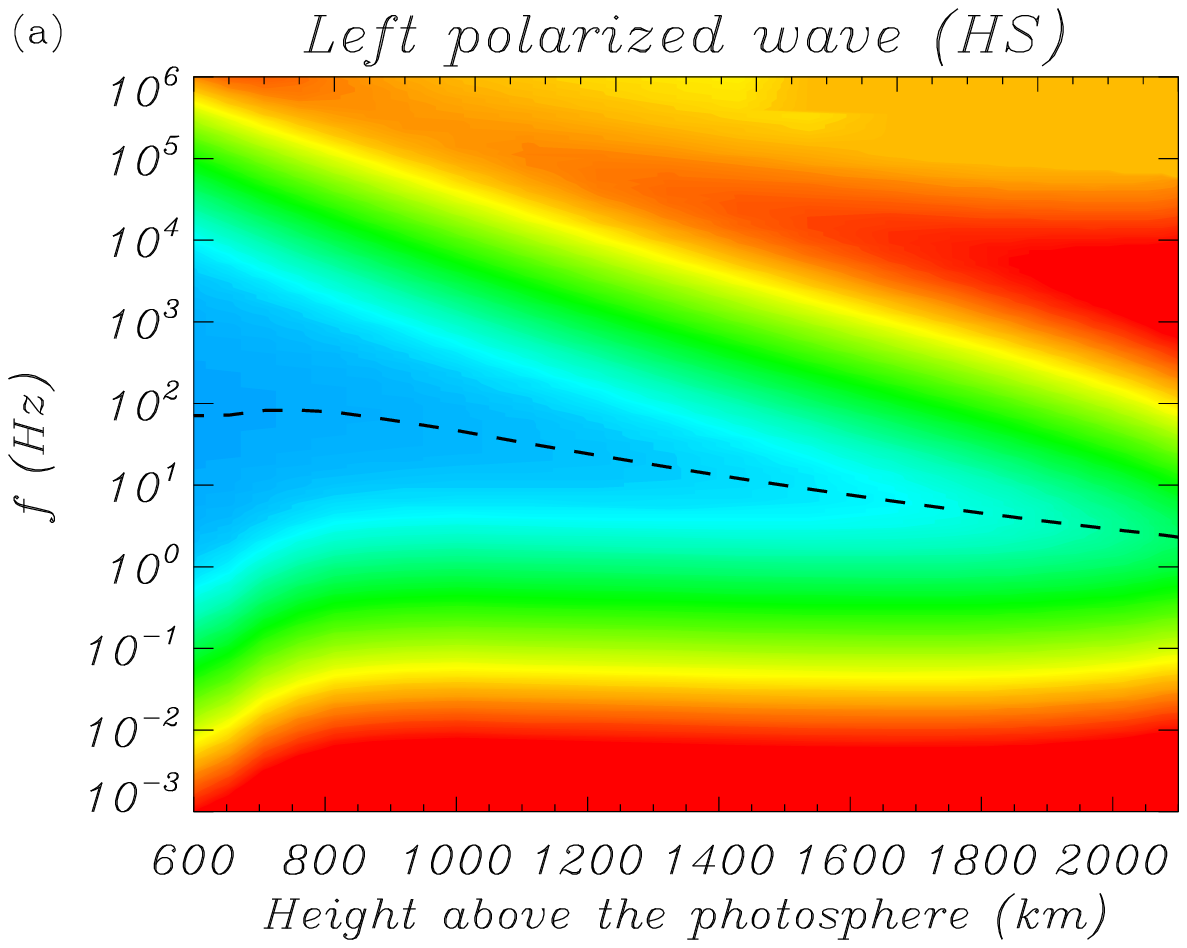}
    \includegraphics[width=.95\columnwidth]{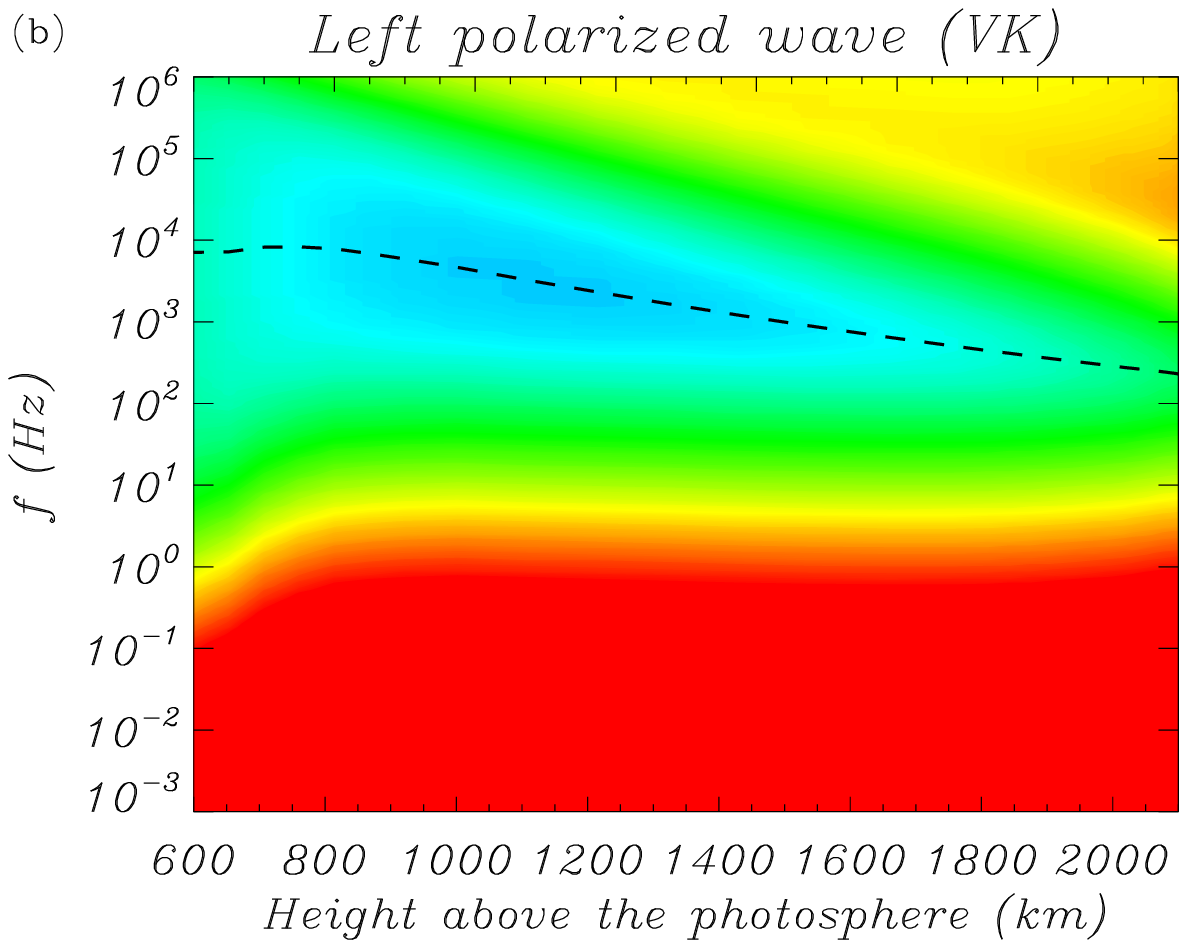}
 \includegraphics[width=.95\columnwidth]{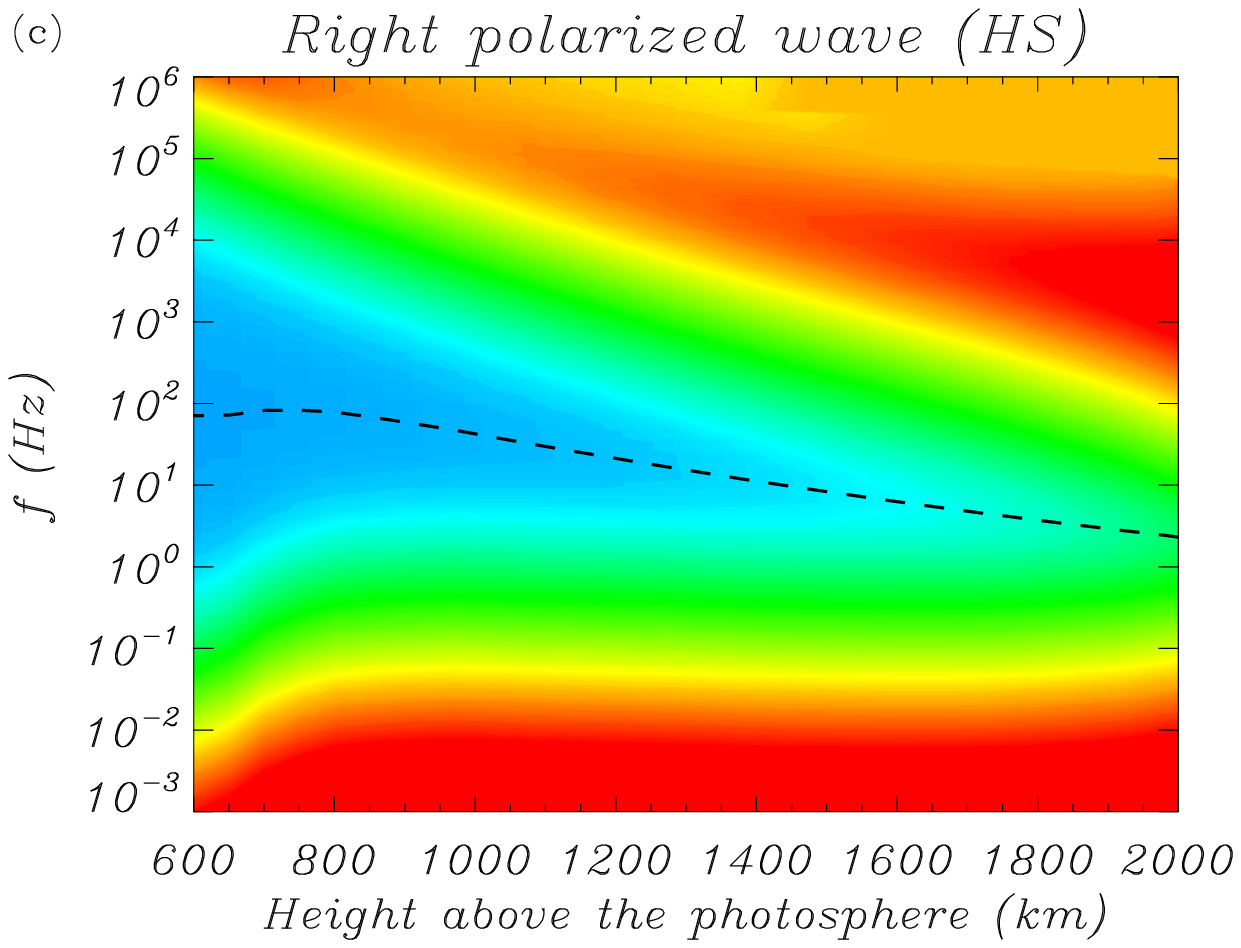}
    \includegraphics[width=.95\columnwidth]{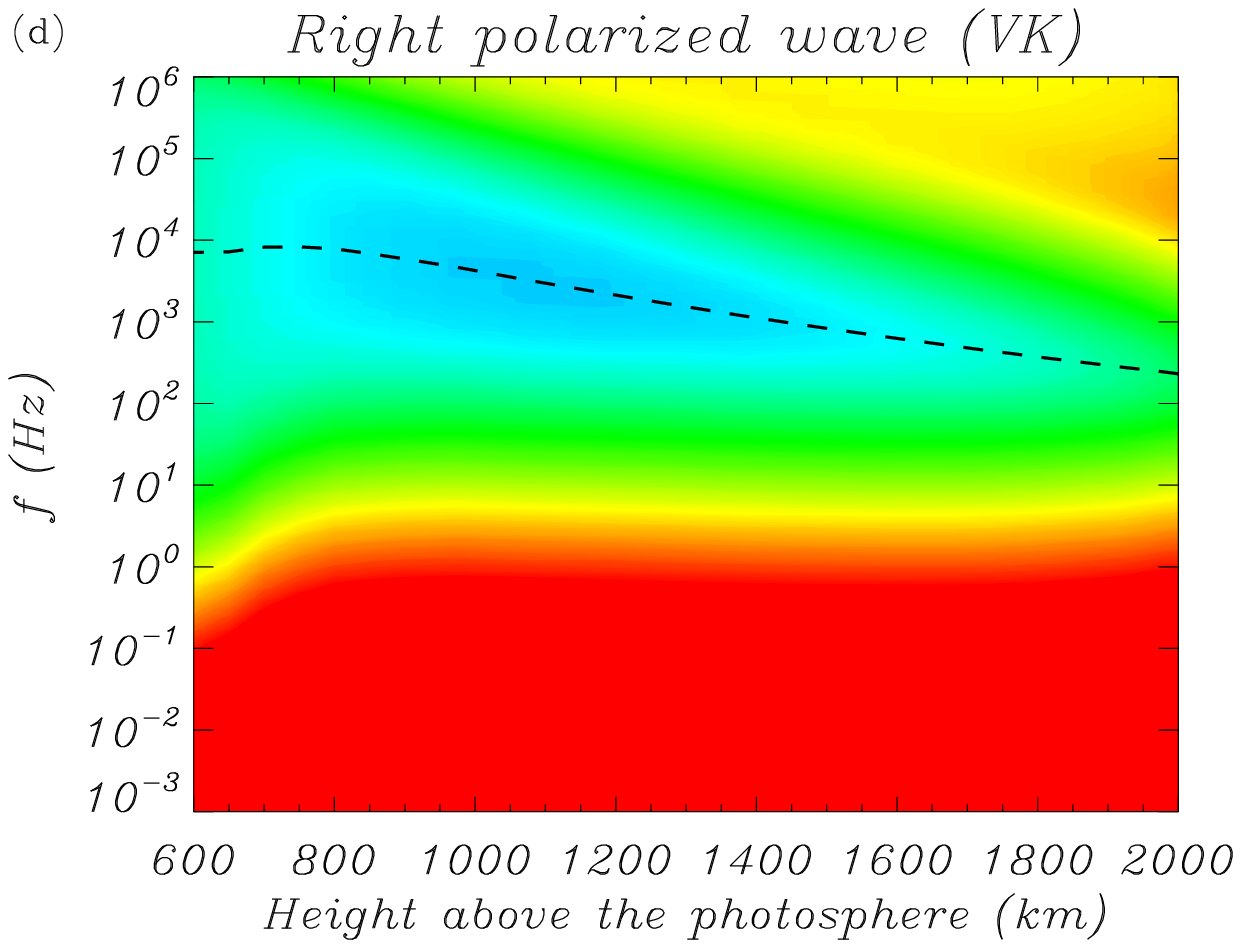}
  \includegraphics[width=1.45\columnwidth]{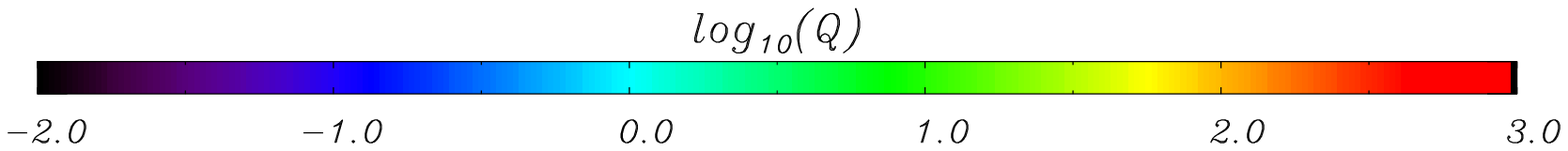}
   \caption{Contour plots of the quality factor $Q$ (in logarithmic scale) of periodically driven Alfv\'en waves in the upper solar chromosphere as function of the  frequency $f$ (vertical axis in logarithmic scale) and height above the photosphere (horizontal axis). Top and botton panels correspond to the left and right polarized waves, respectively.  Left and right panels are the results obtained using the hard-sphere  cross sections (HS) and the \citet{vranjes2013} cross sections (VK), respectively. The black dashed line corresponds to the approximate optimal frequency given in Equation~(\ref{eq:fop}). We note that the VK results are only strictly valid for heights larger than 900~km.}
              \label{fig:spatial1}%
    \end{figure*}

Here we move to study the case of Alfv\'en waves excited by a periodic driver. Thus, we explore the spatial damping of the waves for a fixed frequency. Now the dispersion relation is solved to obtain the complex wavenumber and to compute the quality factor afterwards. The dispersion relation has twelve solutions for the wavenumber. As in the case of the solutions for the frequency explored in the previous subsection, some of the solutions  do not actually correspond to propagating Alfv\'en waves. Instead, these solutions are heavily damped modes because they are directly related to the various dissipative mechanisms included in the main equations. As before, these solutions are discarded from  our study and we only focus on investigating the quality factor of the Alfv\'en waves.

Figure~\ref{fig:spatial1} shows contour plots of the quality factor as a function of height, $h$, above the photosphere for  frequencies, $f=\omega/2\pi$,  in between $10^{-3}$~Hz and $10^6$~Hz. We note that the wavelengths associated with the frequencies of the lower part of this range may be greater than the pressure scale height in the chromosphere. Hence, the local analysis may be compromized for the lowest frequencies considered in Figure~\ref{fig:spatial1}.  Visually, in Figure~\ref{fig:spatial1} we do not see noticeable differences between the left and right polarized solutions. This is so because the considered wave frequencies are lower than the ion cyclotron frequency. A remarkable difference with the case of temporal damping displayed in Figure~\ref{fig:temporal1} is that the  quality factor never goes below $Q=1/2$. In the case of spatial damping the waves are not overdamped. There is, however, a region around a certain optimal frequency where the damping is maximal, i.e., $Q$ reaches its lowest value. This optimal frequency varies with height and depends on the cross sections used in the computations. The optimal frequency is about two orders of magnitude higher in the VK case than in the HS case.  Again, the use of the more accurate VK cross sections has a strong impact on the results.

\begin{figure*}
   \centering
  \includegraphics[width=0.95\columnwidth]{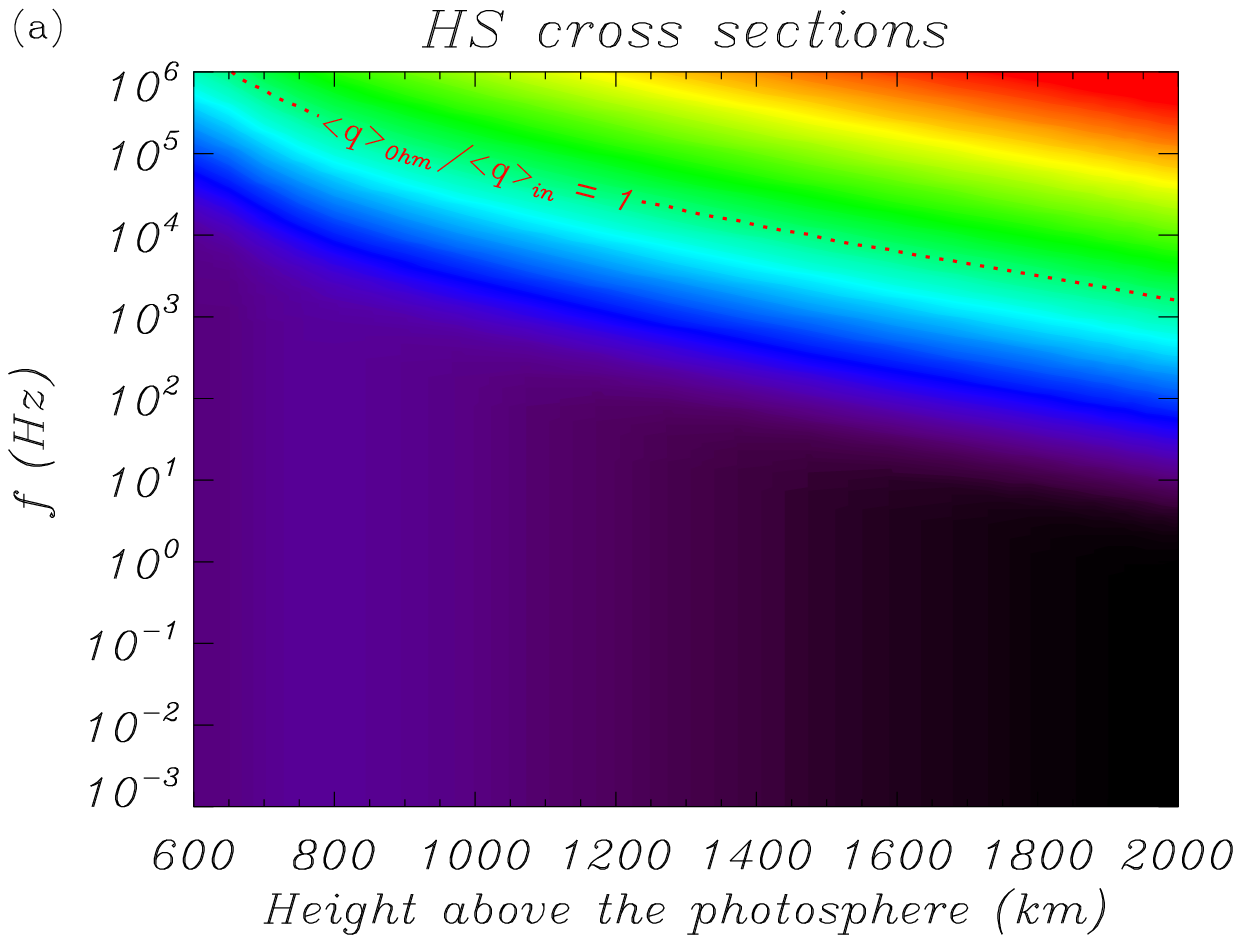}
    \includegraphics[width=0.95\columnwidth]{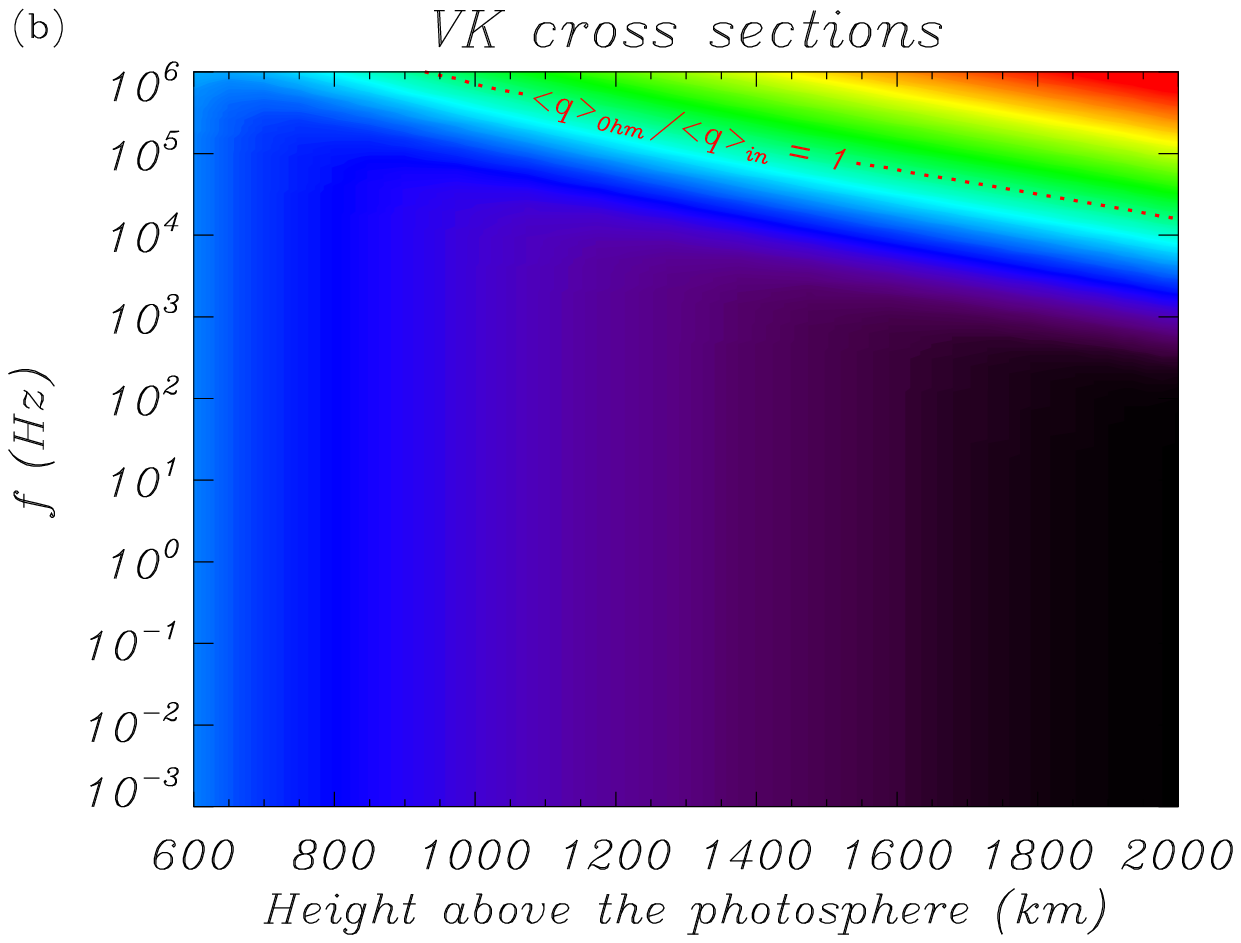}
  \includegraphics[width=1.45\columnwidth]{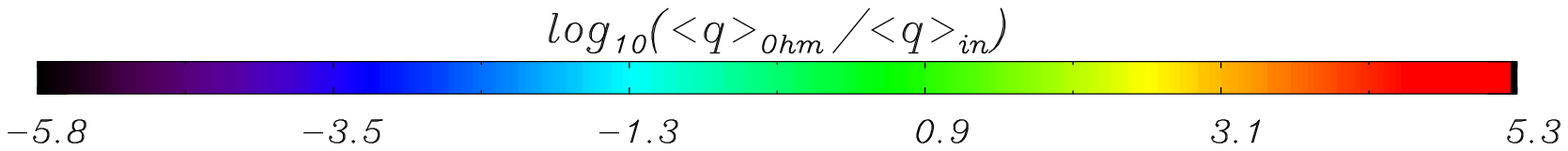}
   \caption{Contour plots of the ratio of Ohm's heating rate to ion-neutral collisions heating rate  (in logarithmic scale) for periodically driven Alfv\'en waves in the upper solar chromosphere as function of the  frequency $f$ (vertical axis in logarithmic scale) and height above the photosphere (horizontal axis). We used the expression derived by  \citet[][Equation~(A5)]{song2011}. Left and right panels are the results obtained using the hard-sphere  cross sections (HS) and the \citet{vranjes2013} cross sections (VK), respectively. The red dotted line denotes the same heating rate for both mechanisms. We note that the VK results are only strictly valid for heights larger than 900~km.}
              \label{fig:heat}%
    \end{figure*}

As before, we can use the previous results by \citet{soler2013} in order to better understand the computations displayed in Figure~\ref{fig:spatial1}. From Equations~(37) and (38) of \citet{soler2013} we can derive  the optimal frequency for the damping of  Alfv\'en waves  due to ion-neutral collisions exclusively. Equations~(37) and (38) of \citet{soler2013} correspond to the real and imaginary parts of the wavenumber, so that an expression of the quality factor can be straightforwardly computed as
\begin{equation}
Q = \sqrt{\frac{1}{4}+\left( \frac{\omega^2 + \left( 1+\chi \right) \nuni^2}{\chi\nuni\omega} \right)^2}.
\end{equation}
We find that the quality factor is minimal, i.e., the damping is maximal, when $\omega = \nuni \sqrt{1+\chi}$, which corresponds to an optimal wave frequency, $f_{\rm op}$, given by
\begin{equation}
f_{\rm op} =  \frac{\nuni}{2\pi}\sqrt{1+\frac{\rhon}{\rhoi}}. \label{eq:fop}
\end{equation}
We overplot in Figure~\ref{fig:spatial1} the optimal frequency computed from Equation~(\ref{eq:fop}) and an excellent agreement is found. In this case, the improved physics taken into account in the present work does not significantly modify the main results by \citet{soler2013} concerning the efficiency of ion-neutral collisions as a damping mechanism for periodically driven Alfv\'en waves.

\section{Discussion}
\label{sec:dis}

In this paper we have investigated theoretically the damping of Alfv\'en waves in the partially ionized chromosphere. The present analysis improves the physical description of the chromospheric plasma considered in previous computations by, e.g., \citet{soler2013}, by including several  processes missing from the previous works.  Ion-neutral collisions have a strong impact on the damping of both impulsively driven and periodically driven Alfv\'en waves in the chromosphere. Thus,  Alfv\'en waves with frequencies near the ion-neutral collision frequency are very efficiently damped. Ion-neutral collisions are the most important damping mechanism in that frequency range, while the other damping mechanisms considered in this work, namely viscosity, magnetic diffusion, and electron-neutral collisions, provide corrections to the damping by ion-neutral collisions. The implications of this strong damping for plasma heating are straightforward.

In the case of waves excited by an impulsive driver, we found that  the strict frequency cut-off due to ion-neutral collisions commonly discussed in the literature \citep[e.g.,][]{kulsrud1969,kamaya1998,soler2013} is here replaced by a critical interval of wavelengths for which Alfv\'en waves are overdamped, i.e., their propagation is inhibited due to the very strong damping. Hall’s current and electron inertia are the main physical effects responsible for removing the strict cut-off \citep[see][]{zaqarashvili2012}. In a recent work, \citet{vranjes2014}  claim that the strict cut-off  occurs regardless of the presence of Hall's current and the electron inertia term. Although \citet{vranjes2014} use a set of equations similar to the one used here and their mathematical analysis  is correct, their conclusions concerning the presence of the strict cut-off do not agree with our findings.  The present results, and those of \citet{zaqarashvili2012}, indicate that strict cut-offs should also be absent from the results of \citet{vranjes2014}.  In connection to the amount of wave energy that can be dissipated in the plasma, the fact that the waves have strict cut-offs ($Q=0$) or are overdamped ($0<Q<1/2$) makes no practical difference. In both cases, all the energy of the disturbance is dissipated {\em in situ} instead of traveling far away as a propagating wave.

All the energy stored in the  impulsively generated perturbations with wavelengths belonging to the critical interval  is necessarily dissipated in the vicinity of the location of the impulsive driver instead of begin transported away by propagating waves. Strong plasma heating might therefore be produced by these overdamped waves. It is seen from Figure~\ref{fig:temporal1} that the impulsively driven waves are overdamped when   the spatial extent of the  perturbation is in the range from 1~km to 50~km in the HS case, and from 1~m to 1~km in the VK case, while the perturbations with spatial extents outside these intervals generate regular propagating waves.  In a realistic chromosphere, where the plasma and the magnetic field dynamically evolve in time, Alfv\'en waves are most likely to be impulsively driven continuously and everywhere in the chromosphere, meaning that overdamped Alfv\'enic perturbations may contribute significantly to chromospheric heating. Another way to excite overdamped, short-wavelength Alfv\'en waves in the chromosphere is by means processes like cascades of energy via magnetohydrodynamic turbulence. However, to determine how the waves are  excited and what is the actual heating rate generated by the overdamped waves, it is necessary, first, to consider self-consistent numerical simulations including the excitation and back reaction of the waves on the plasma and, second, to use the full form of the energy equation taking into account all the possible sources and sinks of energy.

Periodically driven Alfv\'en waves are not overdamped, but there is an optimal driving frequency that produces the strongest damping and thus the strongest energy dissipation. From Figure~\ref{fig:spatial1} we see that the optimal driving frequency varies from 1~Hz to $10^2$~Hz in the HS case, and from $10^2$~Hz to $10^4$~Hz in the VK case. Ion-neutral collisions and Ohmic dissipation are two possible processes that can produce heating for the periodically driven waves. \citet{song2011} derived an expression for the heating rate due to Ohmic dissipation and ion-neutral collisions as a function of the  frequency (see their Equation~(A5)). From the expression derived by \citet{song2011}, the ratio of Ohm's heating rate to ion-neutral collisions heating rate   is
\begin{equation}
\frac{\left< q \right>_{\rm Ohm}}{\left< q \right>_{\rm in}} = \frac{\left( \nuei + \nuen \right)\nuin}{\oci\oce} \left[ \left( \frac{1+\chi}{\chi} \right)^2 + \left( \frac{\omega}{\nuin} \right)^2 \right].  \label{eq:song}
\end{equation}
Equation~(\ref{eq:song}) compares the efficiency of the two heating mechanisms. Figure~\ref{fig:heat} displays the ratio of heating rates computed from Equation~(\ref{eq:song}). Ion-neutral collisions heating is much more important that Ohmic heating in the upper chromosphere unless high driving frequencies are considered. Importantly, ion-neutral collisions heating dominates over Ohmic heating not only for frequencies near the optimal frequency but also for lower frequencies.  Again, there are differences between the results for the HS and the VK cross sections. The VK case requires higher frequencies than the HS case for Ohmic heating to be of importance compared to ion-neutral collisions heating. Viscosity is another dissipative mechanism capable of producing plasma heating. However, we expect  viscous heating to be even less important than Ohmic heating.

Throughout this paper it has been stressed that the value of the ion-neutral cross section plays a very relevant role. This parameter directly determines, among other things, what wavelengths and frequencies are most efficiently damped and how they vary with height in the chromosphere. We found significant differences in the estimations of the critical wavelengths and optimal frequencies depending on the set of cross sections used in the computations. The classic hard-sphere value of the cross section differs about two orders of magnitude from the recent quantum-mechanical calculations of this parameter by \citet{vranjes2013}.  The values proposed by \citet{vranjes2013} take into account several physical effects missing in the hard-sphere model, so that the VK cross sections are more physically accurate than the HS cross sections. Since the numerical value of the cross sections plays an important role, in principle the more accurate VK values are the ones that should be used in theoretical computations.

However,   although the basic atomic physics is obviously the same for laboratory plasmas and the chromospheric plasma, the chromosphere is far from being a controlled environment like a laboratory plasma.  The chromosphere is a multi-thermal, multi-fluid environment. The cross sections measured in controlled laboratory plasmas or those theoretically computed in idealized models, as in \citet{vranjes2013}, may not necessarily match the effective ion-neutral collision cross section in the chromospheric plasma.  The  use of observations of magnetohydrodynamic  waves in the solar atmosphere in combination with the predicted behavior of these waves in theoretical models is an indirect method infer physical properties of the plasma and/or the magnetic field called magnetohydrodynamic seismology \citep[see, e.g.,][]{uchida1970,roberts1984}. We propose that future observations with sufficiently high temporal and spatial resolutions may be able to detect the optimal wavelengths and frequencies for the damping of Alfv\'en waves in the chromosphere. Thus, a seismological test of the effective ion-neutral collision frequency and cross section in the chromospheric plasma may be possible if the required observations become available. In this direction, solar observations performed with  ALMA \citep[see][]{karlicky2011} may shed some light on these important problems and on the role of high-frequency Alfv\'en waves in chromospheric heating.

While in this paper we have investigated Alfv\'en waves, most of the results can be extrapolated to the case of magnetohydrodynamic kink waves, which have a predominantly Alfv\'enic character \citep{goossens2012}. Kink waves have been observed in the chromosphere \citep[e.g.,][]{kukhianidze2006,zaqarashvili2007,okamoto2011,pietarila2011,kuridze2012,morton2014}. Results from \citet{soler2012,soler2013tube} indicate that ion-neutral collisions affect kink waves in a similar manner as they affect pure Alfv\'en waves. The critical wavelengths and optimal frequencies that maximize the damping would be the same for both pure Alfv\'en waves and kink waves. Hence, the main conclusions of this paper are also valid in the case of kink waves.

Finally, we should discuss some of the limitations of this study. The two main simplifications  are, first, the static chromospheric model and, second, the local analysis of the perturbations. On the one hand, the  chromospheric model is based on the FAL93-F static one-dimensional model, which misses part of the highly dynamical behavior observed in the actual chromosphere. On the other hand, the local analysis does not fully capture the behavior of waves with wavelengths similar to or longer than the stratification scale height and ignores their possible reflection \citep{zaqarashvili2013,tu2013}. This second limitation may be overcome in future studies by investigating the propagation and damping of the waves using numerical simulations that consistently include the effect of gravitational stratification as in, e.g., \citet{russell2013} and \citet{tu2013}. Furthermore, in order to properly address the first of the limitations discussed above, these numerical simulations should include a time-dependent dynamic background instead of the static model adopted here. This is an interesting task to be done in forthcoming investigations.

\begin{acknowledgements}
We thank the referee, Dr. Richard Morton, for useful suggestions that helped us to improve the quality of the paper. We acknowledge discussion within the ISSI team on `Partially ionized plasmas in astrophysics' and thank ISSI for their support.  RS and JLB acknowledge support from MINECO and FEDER funds through project AYA2011-22846, and from CAIB through the `Grups Competitius' program and FEDER funds. RS also acknowledges support from MINECO through a `Juan de la Cierva' grant, from MECD through project CEF11-0012, and from the `Vicerectorat d'Investigaci\'o i Postgrau' of the UIB. The work of TZ was supported by the Austrian Fonds zur F\"orderung der Wissenschaftlichen Forschung (project P26181-N27), by the European FP7-PEOPLE-2010-IRSES-269299 project SOLSPANET and by Shota Rustaveli National Science Foundation grant DI/14/6-310/12.
\end{acknowledgements}

\bibliographystyle{aa.bst} 
\bibliography{refs}

\end{document}